\documentclass[sigconf,table,xcdraw,english,bookmarks=false]{acmart}
\usepackage[utf8]{inputenc}

\usepackage{adjustbox}
\usepackage{csvsimple}
\usepackage{multirow}
\usepackage{amsmath}
\usepackage{longtable}
\usepackage{algorithm} 
\usepackage[noend]{algpseudocode}
\usepackage{listings}
\usepackage{tcolorbox}
\usepackage{colortbl}
\usepackage{pgfplots}
\usepackage{pgfplotstable}
\def\code#1{\texttt{#1}}

\pgfplotstableset{
    /color cells/min/.initial=0,
    /color cells/max/.initial=1000,
    /color cells/textcolor/.initial=,
    color cells/.code={%
        \pgfqkeys{/color cells}{#1}%
        \pgfkeysalso{%
            postproc cell content/.code={%
                \begingroup
                \pgfkeysgetvalue{/pgfplots/table/@preprocessed cell content}\value
                \ifx\value\empty
                    \endgroup
                \else
                \pgfmathfloatparsenumber{\value}%
                \pgfmathfloattofixed{\pgfmathresult}%
                \let\value=\pgfmathresult
                \pgfplotscolormapaccess
                    [\pgfkeysvalueof{/color cells/min}:\pgfkeysvalueof{/color cells/max}]
                    {\value}
                    {\pgfkeysvalueof{/pgfplots/colormap name}}%
                \pgfkeysgetvalue{/pgfplots/table/@cell content}\typesetvalue
                \pgfkeysgetvalue{/color cells/textcolor}\textcolorvalue
                \toks0=\expandafter{\typesetvalue}%
                \xdef\temp{%
                    \noexpand\pgfkeysalso{%
                        @cell content={%
                            \noexpand\cellcolor[rgb]{\pgfmathresult}%
                            \noexpand\definecolor{mapped color}{rgb}{\pgfmathresult}%
                            \ifx\textcolorvalue\empty
                            \else
                                \noexpand\color{\textcolorvalue}%
                            \fi
                            \the\toks0 %
                        }%
                    }%
                }%
                \endgroup
                \temp
                \fi
            }%
        }%
    }
}

\usepackage{subcaption}

\usepackage{xcolor}

\definecolor{codegreen}{rgb}{0,0.6,0}
\definecolor{codegray}{rgb}{0.5,0.5,0.5}
\definecolor{codepurple}{rgb}{0.58,0,0.82}
\definecolor{backcolour}{rgb}{0.95,0.95,0.92}
\definecolor{cellgreen}{HTML}{009901}
\definecolor{cellyellow}{HTML}{FFF700}
\definecolor{cellorange}{HTML}{FF6F00}
\definecolor{clgreen}{HTML}{34FF34}
\definecolor{clyellow}{HTML}{FDFF7E}
\definecolor{clorange}{HTML}{FE996B}
\definecolor{cellud}{HTML}{EE442F}
\definecolor{cellue}{HTML}{63ACBE}
\definecolor{celluo}{HTML}{CCBE9F}

\definecolor{cellgray}{HTML}{aeaeae}
 
\lstdefinestyle{mystyle}{
    backgroundcolor=\color{backcolour},   
    commentstyle=\color{codegreen},
    keywordstyle=\color{blue},
    numberstyle=\tiny\color{codegray},
    stringstyle=\color{codepurple},
    basicstyle=\ttfamily\tiny,
    breakatwhitespace=false,         
    breaklines=true,                 
    captionpos=b,                    
    keepspaces=true,                 
    numbers=left,                    
    numbersep=3pt,                  
    showspaces=false,                
    showstringspaces=false,
    showtabs=false,                  
    tabsize=2
}
 
\usepackage{array}
\newcolumntype{P}[1]{>{\centering\arraybackslash}p{#1}}

\title{Defect Prediction Guided Search-Based Software Testing}

\author{Anjana Perera}
\email{Anjana.Perera@monash.edu}
\orcid{0000-0002-5080-9276}
\affiliation{%
  \institution{Faculty of Information Technology}
  \institution{Monash University}
  \city{Melbourne}
  \country{Australia}
}

\author{Aldeida Aleti}
\email{Aldeida.Aleti@monash.edu}
\orcid{0000-0002-1716-690X}
\affiliation{%
  \institution{Faculty of Information Technology}
  \institution{Monash University}
  \city{Melbourne}
  \country{Australia}
}

\author{Marcel B\"{o}hme}
\email{marcel.boehme@acm.org}
\orcid{0000-0002-4470-1824}
\affiliation{%
  \institution{Faculty of Information Technology}
  \institution{Monash University}
  \city{Melbourne}
  \country{Australia}
}

\author{Burak Turhan}
\email{Burak.Turhan@monash.edu}
\orcid{0000-0003-1511-2163}
\affiliation{%
  \institution{Faculty of Information Technology}
  \institution{Monash University}
  \city{Melbourne}
  \country{Australia}
}

\copyrightyear{2020}
\acmYear{2020}
\setcopyright{acmlicensed}\acmConference[ASE '20]{35th IEEE/ACM International Conference on Automated Software Engineering}{September 21--25, 2020}{Virtual Event, Australia}
\acmBooktitle{35th IEEE/ACM International Conference on Automated Software Engineering (ASE '20), September 21--25, 2020, Virtual Event, Australia}
\acmPrice{15.00}
\acmDOI{10.1145/3324884.3416612}
\acmISBN{978-1-4503-6768-4/20/09}

\pgfplotsset{compat=1.16}

\begin{document}
\begin{NoHyper}

\begin{abstract}

Today, most automated test generators, such as search-based software testing (SBST) techniques focus on achieving high code coverage.
However, high code coverage is not sufficient to maximise the number of bugs found, especially when given a limited testing budget. In this paper, we propose an automated test generation technique that is also guided by the estimated degree of defectiveness of the source code. Parts of the code that are likely to be more defective receive more testing
budget than the less defective parts. To measure the degree of defectiveness, we leverage Schwa, a notable defect prediction technique.

We implement our approach into EvoSuite, a state of the art SBST tool for Java. Our experiments on the Defects4J benchmark demonstrate the improved efficiency of defect prediction guided test generation and confirm our hypothesis that spending more time budget on likely defective parts increases the number of bugs found in the same time budget.

\end{abstract}



\maketitle

\section{Introduction}

Software testing is a crucial step in improving software quality. Finding effective test cases, however, is a complex task, which is becoming even more difficult with the increasing size and complexity of software systems. Automated software testing makes this labour intensive task easier for humans by automatically generating test cases for a software system. In particular, search based software testing (SBST) techniques~\cite{harman2015achievements} have been very successful in automatically generating test cases, and are widely used not only in academia, but also in the industry (e.g., Facebook~\citep{alshahwan2018deploying,mao2016sapienz}).

SBST techniques use search methods such as genetic algorithms~\cite{aleti2015test,aleti19testFunc} to find high quality test cases for a particular system. These methods focus on code coverage, and research shows that SBST methods are very effective at achieving high coverage \citep{panichella2017automated, panichella2015reformulating,oliveira2018mapping,aleti2017analysing}. They can even cover more code than the manually written test cases \citep{rueda2015unit,fraser2013does}. However, a test suite with high code coverage does not necessarily imply effective bug detection by the test suite. 
Indeed, previous studies show that SBST methods are not as effective in finding real bugs~\citep{shamshiri2015automatically, almasi2017industrial}.
Even EvoSuite \citep{fraser2011evolutionary} --  a state of the art SBST tool -- could only find on average 23\% of the bugs from the Defects4J dataset \citep{just2014defects4j}, which contains 357 bugs from 5 java projects \citep{shamshiri2015automatically}. Ideally, SBST techniques should aim at generating test cases that reveal bugs, however this is a difficult task since during the search for test cases it is not possible to assess if a test case has found a bug (e.g., semantic bugs). In this paper, we aim to enhance SBST techniques by incorporating information from a defect prediction algorithm to inform the search process of the areas in the software system that are likely to be defective. Thus, the SBST technique, while it cannot tell whether the test cases it produces are indeed finding bugs, it is able to generate more test cases for the defective areas, and as a result, increases the likelihood of finding the bugs.

Defect prediction algorithms \citep{lewis2013does} estimate the likelihood that a file \citep{lewis2013does, de2015software, dam2019lessons}, class \citep{basili1996validation} or method/function \citep{caglayan2015merits, hata2012bug, giger2012method} in a software system is defective. These methods are very effective at identifying the location of bugs in software
\citep{paterson2019empirical, nagappan2008influence, caglayan2015merits}.
As a result of their efficacy, defect prediction models are used to help developers focus their limited testing effort on components that are the most likely to be defective \citep{dam2019lessons}. In addition, defect prediction has been used to inform a test case prioritisation strategy, G-clef~\citep{paterson2019empirical}, of the classes that are likely to be buggy, and found it is promising. Our work is the first to use defect prediction for improving automated test case generation.

We introduce defect prediction guided SBST (SBST$_{DPG}$), which uses information from a defect predictor to focus the search towards the defective areas in software rather than spending the available computational resources (i.e., time budget) to cover non-defective areas. We employ Schwa~\cite{de2015software,paterson2019empirical} as the defect prediction approach, which calculates the defect scores based on the change history of the Java classes. Next, a budget allocation algorithm, called Budget Allocation Based on Defect Scores (BADS) allocates the time budget for each class based on the predictions given by the defect predictor. At a high level, it follows the basic and intuitive rule; highly likely to be defective classes get a higher time budget allocated and less likely to be defective classes get a lower time budget. Finally, we use $DynaMOSA$~\cite{panichella2017automated}, a state of the art SBST algorithm, to generate test suites for each class in the project by spending the allocated time budgets.

Real-world projects are usually very large and there can be even more than 1,000 classes in a project. Hence, it takes significant amount of resources (e.g., time) to run these test generation tools for each class in the project. At the same time, the available computational resources are often limited in practice \citep{campos2014continuous}. Therefore, it is necessary to optimally utilise the available resources (e.g., time budget) to generate test suites for the projects with maximal bug detection. 
Existing SBST approaches allocate the available time budget equally for each class in the project \citep{fraser20151600,fraser2014large}. Usually, most classes are clean, hence we argue that this is a sub-optimal strategy. 
Our proposed approach addresses this by allocating the available time budget to each class in the project based on the class level defect prediction information.

We evaluate how our approach performs in terms of the efficiency in finding real bugs compared to the state of the art. 
Second, we examine if our approach finds more unique bugs. This is particularly important to investigate, as it will reveal if using information from the defect prediction can help SBST reveal bugs that cannot be found otherwise. We evaluate SBST$_{DPG}$ on 434 reported real bugs from 6 open source java projects in the Defects4J dataset. Our empirical evaluation demonstrates that in a resource constrained environment, when given a tight time budget, SBST$_{DPG}$ is significantly more efficient than the state of the art SBST with a large effect size. In particular, SBST$_{DPG}$ finds up to 13.1\% more bugs on average compared to the baseline approach. In addition, our approach is also able to expose more unique bugs which cannot be found by the state-of-the-art approach.

In summary, the contribution of this paper is a novel approach that combines defect prediction and SBST to improve the bug detection capability of SBST by focusing the search more towards the defective areas in software. 
In addition, we present an empirical evaluation involving 434 real bugs from 6 open source java projects (which took roughly 34,600 hours) that demonstrates the efficiency of our proposed solution. Finally, the source code of our proposed technique and the scripts for post processing the results are publicly available here: \url{https://github.com/SBST-DPG}

\section{Related Work}
\label{sec:related_work}

\subsection{Search Based Software Testing}

Search based software testing (SBST) is an effective strategy for achieving high code coverage \citep{panichella2017automated, panichella2015reformulating,panichella2018large}. Shamshiri et al.~\citep{shamshiri2015automatically} and Almasi et al.~\citep{almasi2017industrial} studied the bug detection performance of SBST on open source and industrial software respectively. While EvoSuite \citep{fraser2011evolutionary}, which we consider as the state of the art SBST tool given its maturity, found more bugs than the other techniques used in their studies, overall the results show that the bug detection is still a significant challenge for SBST. Particularly, EvoSuite found only an average of 23\% bugs from the Defects4J dataset \citep{shamshiri2015automatically}. It is clear that using only the 100\% branch coverage criterion was not sufficient to search for test cases that can detect the bugs. In contrast, we use defect prediction information to focus the search to extensively explore the search space for test cases in defective areas.

Gay \cite{gay2017generating} studied the effect of combining coverage criteria on the bug detection performance of SBST, and found that multiple coverage criteria outperform a single criterion. However, the authors did not recommend a general strategy to select which criteria to combine, since their selection strategies also produced many ineffective combinations. Our work is the first approach that focuses on informing SBST of the defective areas to spend more search resources to such areas. Thus, we believe our approach will further improve the bug detection capability of the single criterion or combination of criteria.

\subsection{Defect Prediction}
\label{subsec:relateddp}

Previous work on defect prediction have considered a wide range of metrics such as code size \citep{menzies2006data}, code complexity \citep{zimmermann2007predicting}, object-oriented \citep{basili1996validation}, organisational \citep{nagappan2008influence} and change history \citep{nagappan2005use} to predict future defects in a software project. Graves et al.~\cite{graves2000predicting} showed that the number of changes and particularly the recent changes to the code are effective indicators of future defects. Kim et al.~\citep{kim2007predicting} followed the observation that bugs occur in software change history as bursts, hence they argue that recent changes to the code and recent faults in the code are likely to introduce bugs in the future. 

Rahman et al.~\cite{rahman2011bugcache} proposed a simple approach, which was eventually implemented by the Google Engineering team \citep{googledefect, lewis2013does}, that orders files by the number of bug fix commits in a file, and found out that its performance is quite similar to the more complex approach FixCache \citep{kim2007predicting}. Furthermore, they showed that the files that have been recently involved in a bug fix are likely to contain further bugs. 
Paterson et al.~\cite{paterson2019empirical} used an enhanced version of this approach as the defect predictor to inform a test case prioritisation strategy of the classes that are likely to be buggy, and found it is promising. In particular, they used Schwa \citep{de2015software}, which predicts defects in programs by using three metrics; recent changes, recent bug fixes, and recent new authors \citep{nagappan2008influence} to the code.

\subsection{Budget Allocation Problem}

Search based software test generation tools like EvoSuite generate test suites for each class in the project separately. 
This is done by running a search method such as genetic algorithm (GA) for each class to maximise statement, branch, and method coverage, or a combination of the three. One of the crucial parameters that has to be tuned is the time budget for each class, which is used as a stopping criterion for the GA. Allocating a higher time budget allows the search method to extensively explore the search space of possible test inputs, thus increasing the probability of finding the optimum. 

For small projects, it is feasible to run automated test generation individually for each class in the project. Real-world projects, however, are usually very large, e.g., a modern car has millions of lines of code and thousands of classes \citep{broy2007engineering}, and they require a significant amount of resources (e.g., time) to run the test generation tools for each class in the project. Even in an open source project like Apache Commons Math \citep{commons-math}, there are around 800 classes. In a project like this, it would take at least 13-14 hours to run automated test generation with spending just one minute per each class. At the same time, the available computational resources are often limited in practice \citep{campos2014continuous}. Therefore this raises the question, `How should we optimally utilise the available computational resources (e.g., time budget) to generate test suites for the whole project with maximal bug detection?'.

Previous work on bug detection performance of SBST \citep{almasi2017industrial,shamshiri2015automatically,gay2017generating} allocated a fix time budget to test generation for each buggy class. Since the buggy classes are not known prior to running tests, in practice all the classes in the project have to be allocated the same time budget. Usually, most classes are not buggy, hence we argue that this is a sub-optimal strategy. Our approach solves this problem by allocating time budget to classes based on the information given by a defect predictor.   

Campos et al.~\citep{campos2014continuous} proposed a budget allocation based on the complexity of the classes in order to maximise the branch coverage. In particular, they used number of branches in a class as a proxy to the complexity of the class. In contrast, the scope of this research is to maximise the number of bugs detection.

Contrary to the previous works \citep{almasi2017industrial,shamshiri2015automatically,gay2017generating,campos2014continuous} that considered test suite generation for a regression testing scenario, we focus on generating tests to find bugs not only limited to regressions, but also the bugs that are introduced to the system at various times. 

\begin{figure*}[!ht]
    \centering
    \includegraphics[width=0.8\textwidth]{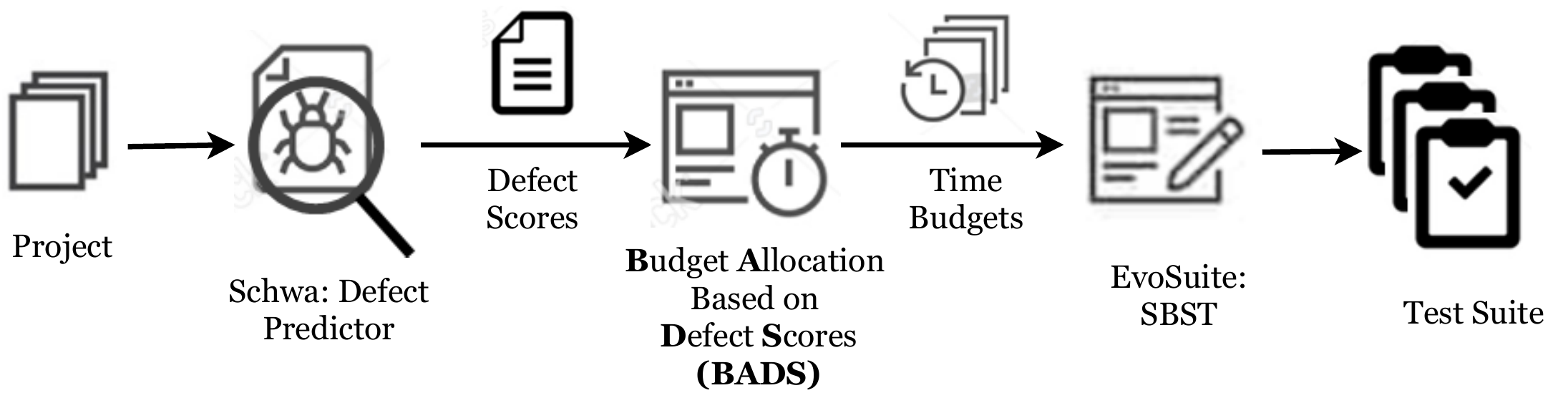}
    \caption{Defect Prediction Guided SBST Overview}
    \label{fig:prioritizer}
\end{figure*}

\section{Defect Prediction Guided Search-Based Software Testing}

Defect Prediction Guided SBST (SBST$_{DPG}$) (see Figure \ref{fig:prioritizer}) uses defect scores of each class produced by a defect predictor to focus the search towards the defective areas of a program. Existing SBST approaches allocate the available time budget equally for each class in the project \citep{fraser2014large,fraser20151600,panichella2017automated,shamshiri2015automatically}. Usually, most classes are not buggy, hence we argue that this is a sub-optimal strategy. Ideally, valuable resources should be spent in testing classes that are likely to be buggy, hence we employ a defect predictor, known as Schwa \citep{de2015software}, to calculate the likelihood that a class in a project is defective. Our approach has three main modules: i) Defect Predictor (DP), ii) \textbf{B}udget \textbf{A}llocation Based on \textbf{D}efect \textbf{S}cores \textbf{(BADS)}, and iii) Search-Based Software Testing (SBST).

\subsection{Defect Predictor}

The defect predictor gives a probability of defectiveness (defect score) for each class in the project. The vector \textbf{s} represents this output. In our implementation of SBST$_{DPG}$, we choose (a) the level of granularity of the defect predictor to be the class level, and (b) the Schwa \cite{de2015software} as the defect predictor module.

Paterson et al.~\cite{paterson2019empirical} successfully applied Schwa as the defect predictor in G-clef to inform a test case prioritisation strategy of the classes that are likely to be buggy. Certainly other defect prediction approaches proposed in the literature (e.g., FixCache~\citep{kim2007predicting}, Change Bursts~\cite{nagappan2010change}) would also be suitable for the task at hand. A strength of Schwa is its simplicity, and that it does not require training a classifier which makes it easy to apply to an industrial setting where training data is not always available (like the one we study). In addition, Schwa can be considered as an enhancement of a tool implemented by the Google Engineering team~\cite{lewis2013does,googledefect}.

Schwa uses the following three measures which have been shown to be effective at producing defect predictions in the literature (see Section \ref{subsec:relateddp}); i) \textit{Revisions} - timestamps of revisions (recent changes are likely to introduce faults), ii) \textit{Fixes} - timestamps of bug fix commits (recent bug fixes are likely to introduce new faults), and iii) \textit{Authors} - timestamps of commits by new authors (recent changes by the new authors are likely to introduce faults). The Schwa tool extracts this information through mining a version control system such as Git \citep{git}. The tool is readily available to use as a python package at Pypi \citep{schwapypi}. 
Therefore, given the robustness of this tool and its approach, we decide to use it as the defect predictor module in our approach.

Schwa \citep{schwagithub} starts with extracting the three metrics; \textit{Revisions} ($R_c$), \textit{Fixes} ($F_c$), and \textit{Authors} ($A_c$) for all classes $c\in C$ in the project. For each timestamp, it calculates a time weighted risk (TWR) \citep{lewis2013does} using the Equation~\eqref{eq:twr}.

\begin{equation}\label{eq:twr}
    TWR(t_{i}) = \frac{1}{1 + \exp(-12t_{i} + 2 + (1 - TR)*10)}    
\end{equation}

The quantity $t_{i}$ is the timestamp normalised between 0 and 1, where 0 is the normalised timestamp of the oldest commit under consideration and 1 is the normalised timestamp of the latest commit. The number of commits that Schwa tracks back in version history of the project ($n$) is a configurable parameter and it can take values from one commit to all the commits.
The parameter $TR \in [0,1]$ is called the Time Range and it allows to change the importance given to the older commits. The time weighted risk formula scores recent timestamps higher than the older ones (see Figure \ref{fig:twr}).

\begin{figure}[!ht]
    \centering
    \includegraphics[width=0.30\textwidth]{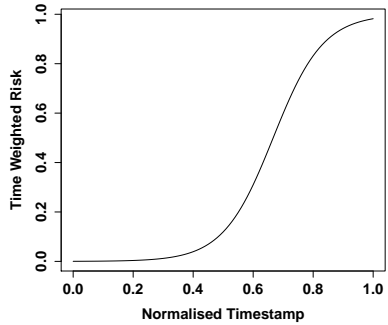}
    \caption{Time Weighted Risk ($TR$ = 0.4)}
    \label{fig:twr}
\end{figure}

Once Schwa calculated the TWRs, it aggregates these TWRs per each metric, and calculates a weighted sum $s_{c}$ for each class $c\in C$ in the project as in Equation~\eqref{eq:schwascore}.

\begin{equation}\label{eq:schwascore}
\begin{split}
    s_{c} & = w_{r}*\sum\limits_{t_{i} \in R_c}TWR(t_{i}) + w_{f}*\sum\limits_{t_{i} \in F_c}TWR(t_{i})\\& \quad + w_{a}*\sum\limits_{t_{i} \in A_c}TWR(t_{i})
\end{split}
\end{equation}

The sum $\sum_{t_{i} \in R_c} TWR(t_{i})$ is the total of the time weighted risks of the \textit{Revisions} metric for class $c$. Similarly, $\sum_{t_{i} \in F_c} TWR(t_{i})$ and $\sum_{t_{i} \in A_c} TWR(t_{i})$ are the sums of the TWRs of the \textit{Fixes} and \textit{Authors} metrics for class $c\in C$. The quantities $w_{r}$, $w_{f}$, and $w_{a}$ are weights that modify the TWR sum of each metric and their sum is equal to 1. The weighted sum, $s_c$, is called the score of class $c\in C$. 

Finally, the Schwa tool estimates the probability $p(c)$ of that a class $c$ is defective as given in Equation~\eqref{eq:schwaprob}.

\begin{equation}\label{eq:schwaprob}
    p(c) = 1 - \exp(-s_c)
\end{equation}

In this paper, we refer to this probability of defectiveness  $p(c)$ as the defect score of class $c\in C$.

\subsection{Budget Allocation Based on Defect Scores}

Budget Allocation Based on Defect Scores (BADS) takes the defect scores ($\mathbf{s}=\{p(c) | c\in C\}$) as input and decides on how to allocate the available time budget to each class based on these scores, producing a vector \textbf{t} as output. Ideally, all the defective classes in the project should get more time budget while non-defective classes can be left out from test generation. However, the defect predictor only gives an estimation of the probability of defectiveness. Therefore, BADS allocates more time budget to the highly likely to be defective classes than to the less likely to be defective classes. This way we expect SBST to get higher time budget to extensively explore for test cases in defective classes rather than in non-defective ones.

\subsubsection{Exponential Time Budget Allocation Based on Defect Scores}
\label{subsec:exp_time_BADS}

\begin{algorithm}[!ht]
\caption{Exponential Time Budget Allocation Based on Defect Scores}\label{algo:expowdrank}
\begin{flushleft}
\textbf{Input:  } The set of all the classes $C$, where $N = |C|$\\
\hspace*{\algorithmicindent} \hspace*{\algorithmicindent}  \textbf{s} = \{$s_{1}, s_2, \ldots , s_N$\}\\
\hspace*{\algorithmicindent} \hspace*{\algorithmicindent} $T, t_{\min}, T_{DP}$\\
\hspace*{\algorithmicindent} \hspace*{\algorithmicindent} $e_{a}, e_{b}, e_{c}$\\
\textbf{Output:} \textbf{t} = \{$t_{1}, t_2, \ldots , t_N$\}\\
\end{flushleft}
\begin{algorithmic}[1]
\Procedure{AllocateTimeBudget}{}
\State \textbf{r} $\gets$ \Call{Assign-Rank}{\textbf{s}}
\State \textbf{r$^{'}$} $\gets$ \Call{Normalise-Rank}{\textbf{r}}
\State \textbf{w$^{'}$} $\gets \emptyset$ 
\For{\textbf{\textit{all}} $c_{i} \in C$}
    \State $w'_{i} \gets e_{a} + e_{b} * \exp(e_{c} * r'_{i})$ \label{algoline:exp} 
\EndFor{}
\State \textbf{w} $\gets$ \Call{Normalise-Weight}{\textbf{w$^{'}$}}
\State \textbf{t} $\gets \emptyset$ 
\For{\textbf{\textit{all}} $c_{i} \in C$}
    \State $t_{i} \gets w_{i} * (T - N * t_{\min} - T_{DP}) + t_{\min}$ \label{algoline:timealloc} 
\EndFor{}
\State \Call{Return}{\textbf{t}}
\EndProcedure
\end{algorithmic}
\end{algorithm}

Algorithm \ref{algo:expowdrank} illustrates the proposed time budget allocation algorithm of BADS, where \textbf{s} is the set of defect scores of the classes, $T$ is the total time budget for the project, $t_{\min}$ is the minimum time budget to be allocated for each class, $T_{DP}$ is the time spent by the defect predictor module, and $e_a$, $e_b$, and $e_c$ are parameters of the exponential function that define the shape of the exponential curve. \textbf{t} is the set of time budgets allocated for the classes.

The defect scores assignment in Figure \ref{fig:defectscores_chart9} is a good example of the usual defect score distribution by a defect predictor. Usually, there are only a few classes which are actually buggy. Allocating higher time budgets for these classes would help maximise the bug detection of the test generation tool. Following this observation and the results of our pilot runs, we use an exponential function (line~\ref{algoline:exp} in Algorithm~\ref{algo:expowdrank}) to highly favour the budget allocation for the few highly likely to be defective classes.

Moreover, there is relatively higher number of classes which are moderately likely to be defective (e.g., 0.5 < defect score < 0.8). It is also important to ensure there is sufficient time budget allocated for these classes. Otherwise, neglecting test generation for these classes could negatively affect bug detection of the test generation tool. We introduce a minimum time budget, $t_{\min}$, to all the classes because we want to ensure that every class gets a budget allocated regardless of the defectiveness predicted by the defect predictor. The exponential function in Algorithm \ref{algo:expowdrank} together with $t_{\min}$ allow an adequate time budget allocation for the moderately likely to be defective classes.

\begin{figure}[!ht]
    \centering
    \includegraphics[width=0.32\textwidth]{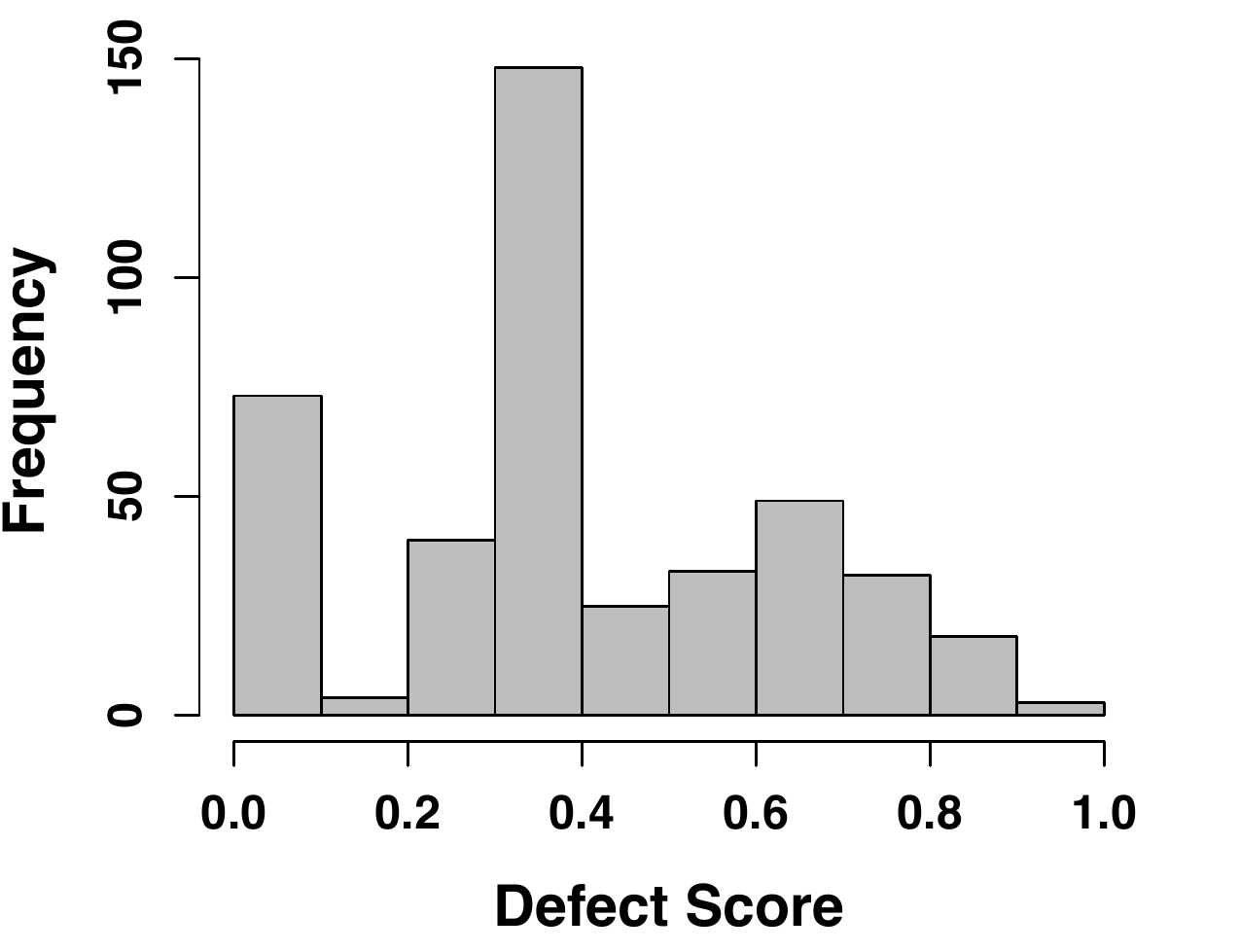}
    \caption{Distribution of the defect scores assigned by Schwa for the classes in Chart-9 bug from Defects4J.}
    \label{fig:defectscores_chart9}
\end{figure}

Upon receiving the defect scores (\textbf{s}), BADS assigns ranks (\textbf{r}) for all the classes according to the defect scores. Next, the \textsc{Normalise-Rank} function normalises the ranks in the range [0,1], where the rank of the most likely to be defective class is 0 and the least likely to be defective class is 1. Then, each class gets a weight ($w^{'}_{i}$) assigned based on its normalised rank by the exponential function. The amount of time budget allocated to class $c_{i}$ is proportional to $w^{'}_{i}$. 
The parameters $e_a$, $e_b$, and $e_c$ have to be carefully selected such that the weights are almost equal and significantly small for the lower-ranked classes, and the difference between the weights of adjacently ranked classes rapidly increases towards the highly-ranked classes. The \textsc{Normalise-Weight} function normalises the weights to the range [0,1], ensuring the summation is equal to 1, and produces the normalised weights vector \textbf{w}. Finally, BADS allocates time budget for each class from the remaining available time budget, $T - N * t_{\min} - T_{DP}$, based on its normalised weight (line~\ref{algoline:timealloc} in Algorithm \ref{algo:expowdrank}) . 

\subsubsection{The 2-Tier Approach}
\label{subsec:2tier}

According to the defect predictor outcome, almost all the classes in the project get non-zero defect scores attached to them. This gives the impression that all these classes can be defective with at least a slight probability. However, in reality, this does not hold true. For a given project version, there are only a few defective classes. A defect predictor is likely to predict that clean classes are also defective with a non-zero probability. While the exponential function disfavours the budget allocation for these less likely to be defective classes, $t_{\min}$ guarantees a minimum time budget allocated to them. If we decrease $t_{\min}$ in order to make the budget allocation negligible for the likely to be clean classes, then it would risk a sufficient time budget allocation for the moderately likely to be defective classes.

We propose the 2-Tier approach which divides the project into two tiers following the intuition that only a set of classes are defective in a project. 
BADS sorts the classes into two tiers before the weights assignment, such that the highly likely to be defective classes are in the \textit{first tier} and the less likely to be defective classes are in the \textit{second tier}. This allows to further discriminate the less likely to be defective classes, and favour the highly likely to be defective classes by simply allocating only a smaller fraction of the total time budget to the \textit{second tier} and allocating the rest to the \textit{first tier}. Section \ref{subsec:param_settings_BADS} provides more details on the parameter selection of the 2-Tier approach.

\subsection{Search Based Software Testing}
\label{subsec:sbst}

We use EvoSuite \citep{fraser2011evolutionary} as the search-based software testing (SBST) module in our defect prediction guided SBST approach. EvoSuite is an automated test generation framework that generates JUnit test suites for Java classes. It was first proposed by Fraser and Arcuri~\cite{fraser2011evolutionary} in 2011, since then it has gained growing interest in the SBST community \citep{rojas2017detailed, panichella2015reformulating, panichella2017automated, arcuri2016unit, shamshiri2015automatically}. Its effectiveness has been evaluated on open source and as well as industrial software projects in terms of the code coverage \citep{panichella2017automated, panichella2015reformulating, rojas2017detailed, panichella2018large, fraser2014large} and bug detection \citep{shamshiri2015automatically,almasi2017industrial}. Furthermore, EvoSuite won 6 out of 7 of the SBST unit testing tool competitions \citep{campos2019evosuite,fraser2016evosuite,fraser2013SBSTtool,fraser2014SBSTtool,fraser2018sbsttool,fraser2017sbsttool}. To date, EvoSuite is actively maintained, and its source code and releases are readily available to use at GitHub \citep{evosuitegithub} and their website \citep{evosuiteweb}. Therefore, given the maturity of EvoSuite, we decide to use it as the SBST module in our approach.

More recently, Panichella et al.~\cite{panichella2017automated} developed a new search algorithm, $DynaMOSA$ (\textit{Dynamic Many-Objective Sorting Algorithm}), as an extension to EvoSuite, which stands as the current state of the art. It has been shown to be effective at achieving high branch, statement and strong mutation coverage than the previous versions of EvoSuite (\citep{panichella2015reformulating,rojas2017detailed,fraser2012whole}) \citep{panichella2017automated}. Moreover, $DynaMOSA$ was the search algorithm of EvoSuite, which won the unit testing tool competition at SBST 2019 \citep{campos2019evosuite}. Therefore, we use $DynaMOSA$ as the search algorithm in EvoSuite.

\section{Design of Experiments}\label{sec:designExp}
We evaluate our approach in terms of its efficiency in finding bugs, and the effectiveness in revealing unique bugs, i.e., bugs that cannot be found by the benchmark approach. Our first research question is:

\begin{center}
    \textit{RQ1: Is SBST$_{DPG}$ more efficient in finding bugs compared to the state of the art?}
\end{center}
To answer this research question, we run a set of experiments where we compare our approach against the baseline method discussed in Section~\ref{subsec:baseline}. All methods are employed to generate test cases for Defects4J~\cite{defects4jweb}, which is a well-studied benchmark of buggy programs described in Section~\ref{defects4J}. Once the test cases are generated, we check if they find the bugs in the programs, and report the results as the mean and median over 20 runs. To check for statistical significance of the differences and the effect size, we employ two-tailed non-parametric Mann-Whitney U-Test with the significance level ($\alpha$) 0.05 \citep{arcuri2014hitchhiker} and Vargha and Delaney's $\widehat{A}_{12}$ statistic \citep{vargha2000critique}. We also plot the results as boxplots to visualise their distribution.

In addition, to analyse the effectiveness of the proposed approach, we seek to answer the following research question:

\begin{center}
    \textit{RQ2: Does SBST$_{DPG}$ find more unique bugs?}
\end{center}

To answer this research question, we analyse the results from the experiments in more detail. While the first research question focuses on the overall efficiency, in the second research question we aim to understand if SBST$_{DPG}$ is capable of revealing more unique bugs which can not be exposed by the baseline method. Part of the efficiency of our proposed method, however, could be due to its robustness, which is measured by the success rate, hence we also report how often a bug is found over 20 runs. 

\subsection{Time Budget}

In real world scenario, total time budget reserved for test generation for a project depends on how it is used in the industry. For example, a project having hundreds of classes and running SBST 1-2 minutes per class takes several hours to finish test generation. If an organisation wants to adapt SBST in their continuous integration (CI) system \citep{fowler2006continuous}, then it has to share the resources and schedules with the processes already in the system; regression testing, code quality checks, project builds etc. In such case, it is important that SBST uses minimal resources possible, such that it does not idle other jobs in the system due to resource limitations.

Panichella et al.~\cite{panichella2017automated} showed that \textit{DynaMOSA} is capable of converging to the final branch coverage quickly, sometimes with a lower time budget like 20 seconds. This is particularly important since faster test generation allows more frequent runs and thereby it makes SBST suitable to fit into the CI/CD pipeline. Therefore, we decide that 30 seconds per class is an adequate time budget for test generation and 15 seconds per class is a tight time budget in a usual resource constrained environment. We conduct experiments for 2 cases of total time budgets ($T$); $15*N$ and $30*N$ seconds.

\subsection{Experimental Subjects}\label{defects4J}

We use the Defects4J dataset \citep{just2014defects4j, defects4jweb} as our benchmark. It contains 434 real bugs from 6 real-world open source Java projects. We remove 4 bugs from the original dataset~\citep{defects4jweb} since they are not reproducible under Java 8, which is required by EvoSuite.
The projects are JFreeChart (26 bugs), Closure Compiler (174 bugs), Apache commons-lang (64 bugs), Apache commons-math (106 bugs), Mockito (38 bugs) and Joda-Time (26 bugs). For each bug, the Defects4J benchmark gives a buggy version and a fixed version of the program. The difference between these two versions of the program is the applied patch to fix the bug, which indicates the location of the bug. The Defects4J benchmark also provides a high-level interface to perform tasks like running the generated tests against the other version of the program (buggy/fixed) to check if the tests are able to find the bug, fixing the flaky test suites etc.~\cite{defects4jweb}. 

Defects4J is widely used for research on automated unit test generation \citep{shamshiri2015automatically}, automated program repair \citep{le2016history}, fault localisation \citep{pearson2017evaluating}, test case prioritisation \citep{paterson2019empirical}, etc. This makes Defects4J a suitable benchmark for evaluating our approach, as it allows us to compare our results to existing work. 

\subsection{Baseline Selection}
\label{subsec:baseline}

We use the current state of the art SBST algorithm, $DynaMOSA$~\cite{panichella2017automated}, with equal time budget allocation, SBST$_{noDPG}$, as our baseline for comparison. Previous work on bug detection capability of SBST allocated an equal time budget for all the classes \citep{shamshiri2015automatically,almasi2017industrial,gay2017generating}. Even though, Campos et al.~\cite{campos2014continuous} proposed a budget allocation targeting the maximum branch coverage, we do not consider this as a baseline in our work as we focus on bug detection instead. Our intended application scenario is generating tests to find bugs not only limited to regressions, but also the bugs that are introduced to the system in different times. Hence, we consider generating tests to all of the classes in the project regardless of whether they have been changed or not. Therefore, in equal budget allocation, total time budget is equally allocated to all the classes in a project.

\subsection{Parameter Settings}

There are 3 modules in our approach. Each module has various parameters to be configured, and the following subsections outline the parameters and their chosen values in our experiments.

\subsubsection{Schwa}

Schwa has 5 parameters to be configured; $w_r$, $w_f$, $w_a$, $TR$, and $n$. We choose the default parameter values used in Schwa \citep{schwagithub} as follows: $w_r = 0.25$, $w_f = 0.5$, $w_a = 0.25$, and $TR = 0.4$. Our preliminary experiments with $n = 50, 100, 500, 1000$ and $all\; commits$ suggest that $n = 500$ gives most accurate predictions.

\subsubsection{EvoSuite}
\label{sec:param_settings_evosuite}

Arcuri and Fraser~\cite{arcuri2013parameter} showed that parameter tuning of search algorithms is an expensive and long process, and the default values give reasonable results when compared to tuned parameters. Therefore, we use the default parameter values used in EvoSuite in previous work \citep{panichella2017automated, fraser2012whole} except for the following parameters.

\textit{Coverage criteria:} We use branch coverage since it performs better among the other single criteria \citep{gay2017generating}. Gay~\cite{gay2017generating} found some multiple criteria combinations to be effective on bug detection than single criterion. However, they did not recommend a strategy to combine multiple criteria as their strategies also produced ineffective combinations. Therefore, we decide to use only single criterion. 

\textit{Assertion strategy:} As Shamshiri et al.~\cite{shamshiri2015automatically} mentioned, mutation-based assertion filtering can be computationally expensive and lead to timeouts sometimes. Therefore, we use all possible assertions as the assertion strategy.

Given a coverage criterion (e.g., branch coverage), $DynaMOSA$ explores the search space of possible test inputs until it finds test cases that cover all of the targets (e.g., branches) or the time runs out (i.e., time budget). These are known as stopping criteria. This way, if the search achieves 100\% coverage before the timeout, any remaining time budget will be wasted. At the same time, $DynaMOSA$ aims at generating only one test case to cover each target in the system under test (SUT), since its objective is to maximise the coverage criterion given. This also helps in minimising the test suite produced. However, when it comes to finding bugs in the SUT, just covering the bug does not necessarily imply that the particular test case can discover the bug. Hence, we find that using 100\% coverage as a stopping criterion and aiming at finding only one test case for each target deteriorate the bug detection capability of $DynaMOSA$. Therefore, in our approach, we configure $DynaMOSA$ to generate more than one test case for each target in the SUT, retain all these test cases, disable test suite minimisation and remove 100\% coverage from the stopping criteria. By doing this, we compromise the test suite size in order to increase the bug detection capability of SBST.

\subsubsection{BADS}
\label{subsec:param_settings_BADS}

Following the results of our pilot runs, we use the default threshold of 0.5 to allocate the classes into the two tiers. In particular, the top half of the classes (ranked in descending order according to defect scores) are allocated in the \textit{first tier} ($N_1$) and the rest are in the \textit{second tier} ($N_2$). $N_1$ and $N_2$ are the number of classes in the \textit{first} and \textit{second tiers} respectively. 

Our preliminary results also suggest that allocating 90\% and 10\% of the total time budget ($T$) to the \textit{first tier} ($T_1$) and the \textit{second tier} ($T_2$) sufficiently favours the highly likely to be defective classes, while not leaving out the less likely to be defective classes from test generation. In particular, we choose $T_1=27*N_1$ and $T_2=3*N_2$ seconds at $T=15*N$ and $T_1=54*N_1$ and $T_2=6*N_2$ seconds at $T=30*N$. We choose 15 and 30 seconds as $t_{min}$ for the \textit{first tier} ($t_{\min_1}$) at $T=15*N$ and $T=30*N$ respectively. The rationale behind choosing these values for $t_{\min_1}$ is that it guarantees the classes in the \textit{first tier} at least get a time budget of the equal budget allocation (i.e., budget allocation without defect prediction guidance). For $t_{min}$ of the \textit{second tier} ($t_{\min_2}$), we assign 3 and 6 seconds at $T=15*N$ and $T=30*N$ because we believe $T_2$ is not enough to go for an exponential allocation.

The parameters for the exponential function are as follows: $e_a = 0.02393705$, $e_b = 0.9731946$, and $e_c = -10.47408$. The rationale behind choosing the parameter values for the exponential function is as follows. The exponential curve is almost flat and equal to 0 for the values in the $x$ axis from 0.5 to 1 (see Figure \ref{fig:exp_func}). Then, after $x = 0.5$, it starts increasing towards $x = 0$. Finally, at $x = 0$, the output is equal to 1.

\begin{figure}[!ht]
    \centering
    \includegraphics[width=0.30\textwidth]{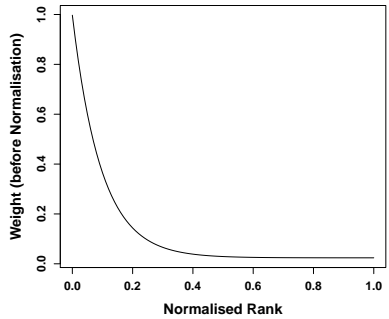}
    \caption{Exponential Function of BADS. $e_a = 0.02393705$, $e_b = 0.9731946$, and $e_c = -10.47408$}
    \label{fig:exp_func}
\end{figure}

\subsection{Prototype}

We implement the Defect Prediction Guided SBST approach in a prototype tool in order to empirically evaluate it. The prototyped tool is available to download from here: \url{https://github.com/SBST-DPG/sbst-dpg}

\subsection{Experimental Protocol}
\label{subsec:experiment_protocol}

As we mentioned earlier, to answer \textit{RQ1} and \textit{RQ2}, we conduct experiments for $T = 15*N$ and $30*N$ seconds.

In SBST$_{DPG}$, Schwa uses current versions of the repositories of the projects. For each bug, Schwa predicts the defectiveness of the classes at the commit just before the bug fixing commit.
For each bug in Defects4J, there is a buggy version and a fixed version of the project. We take each buggy version of the projects, and then generate test suites only for the buggy class(es) of that project version using the two approaches. To take the randomness of SBST into account, we repeat each test generation run 20 times, and carry out statistical tests when necessary. Consequently, we have to run a total of 2 (approaches) $*$ 511 (buggy classes) $*$ 20 (repetitions) $*$ 2 (time budgets) $=$ 40,880 test generations. We collect the generated test suites after each test generation run. Next, we use the \textit{fix test suite} interface in Defects4J to remove the flaky tests from each test suite \citep{shamshiri2015automatically}. Then, we execute each resulting test suite against the respective buggy and fixed versions to check if it finds the bug or not using the \textit{run bug detection} interface. If the test suite is not compilable or there is at least one failing test case when the test suite is run against the buggy version, then it is marked as \textit{Broken}. If not, it will be run against the fixed version. Then, if at least one test case fails, the test suite is marked as \textit{Fail} (i.e., test suite finds the bug). If all the test cases pass, then the test suite is marked as \textit{Pass} (i.e., test suite does not find the bug).

\section{Results}

We present the results for each research question following the method described in Section~\ref{sec:designExp}. While the main aim is to evaluate if our approach is more efficient than the state of the art, we also focus on explaining its strengths and weaknesses.

\begin{table}[h]
\renewcommand{\arraystretch}{1.1}
\caption{Mean and median number of bugs found by the 2 approaches against different total time budgets.}
\label{table:rq1_results_1}
\centering
\begin{adjustbox}{max width=0.48\textwidth}
\begin{tabular}{c|c|c|c|c|c|c}
\hline
 & \multicolumn{2}{c|}{Mean}                    & \multicolumn{2}{c|}{Median}   & \multirow{2}{*}{p-value}                                                                             & \multirow{2}{*}{$\widehat{A}_{12}$}                                                               \\ \cline{2-5}
\multirow{-2}{*}{T (s)}               & SBST$_{DPG}$    & SBST$_{noDPG}$ & SBST$_{DPG}$   & SBST$_{noDPG}$  &   &  \\ \hline
$15*N$             & \textbf{151.45} & 133.95              & \textbf{150.5} & 134.0                 & \textbf{\textless 0.0001}                                                    & \textbf{0.94}                                                                 \\ 
$30*N$  & \textbf{171.45}  & 166.9                   & \textbf{170} & 167.5                  & 0.0671                                                                                          & 0.67                                                                                 \\ \hline
\end{tabular}
\end{adjustbox}
\end{table}

\subsection*{RQ1. Is SBST$_{DPG}$ efficient in finding bugs?}

As described in Section~\ref{sec:designExp}, we perform 20 runs for each SBST approach and each buggy program in Defects4J and report the results as boxplots in Figure~\ref{fig:rq1_results_1}. As we can see, overall, our proposed method SBST$_{DPG}$ finds more bugs than the baseline approach for both 15 and 30 seconds time budgets. 

\begin{figure}[!ht]
    \centering
    \includegraphics[width=0.46\textwidth]{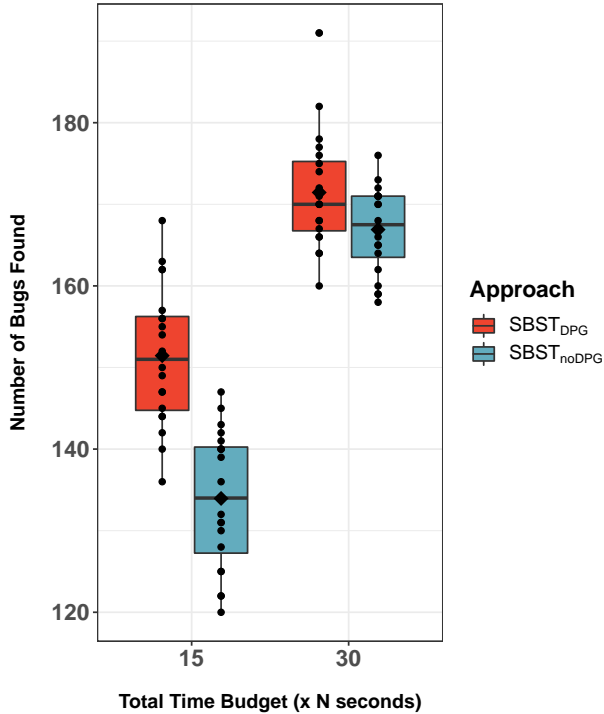}
    \caption{The number of bugs found by the 2 approaches against different total time budgets}
    \label{fig:rq1_results_1}
\end{figure}

We also report the means, medians and the results from the statistical analysis in Table \ref{table:rq1_results_1}. SBST$_{noDPG}$ finds 133.95 bugs on average at total time budget of $15$ seconds per class. SBST$_{DPG}$ outperforms SBST$_{noDPG}$, and finds 151.45 bugs on average, which is an average improvement of 17.5 (+13.1\%) more bugs than SBST$_{noDPG}$. The difference of the number of bugs found by SBST$_{DPG}$ and SBST$_{noDPG}$ is statistically significant according to the Mann-Whitney U-Test (p-value < 0.0001) with a large effect size ($\widehat{A}_{12}$ = 0.94). Thus, we can conclude that SBST$_{DPG}$ is more efficient than SBST$_{noDPG}$. 

At total time budget of $30$ seconds per class, SBST$_{DPG}$ finds more bugs than the SBST$_{noDPG}$. According to the Mann-Whitney U-Test, the difference between SBST$_{DPG}$ and SBST$_{noDPG}$ is not statistically significant, with a p-value of 0.067. However the effect size of 0.67 suggests that SBST$_{DPG}$ finds more bugs than SBST$_{noDPG}$ 67\% of the time, which is significant given how difficult it is to find failing test cases \citep{habib2018many}.

\begin{tcolorbox}
In summary, defect prediction guided SBST (SBST$_{DPG}$) is significantly more efficient than SBST without defect prediction guidance (SBST$_{noDPG}$) when they are given a tight time budget in a usual resource constrained scenario. When there is sufficient time budget SBST$_{DPG}$ is more effective than SBST$_{noDPG}$ 67\% of the time.
\end{tcolorbox}

\begin{figure}[htbp]
\renewcommand{\arraystretch}{1.2}
    \centering
    \includegraphics[width=\linewidth]{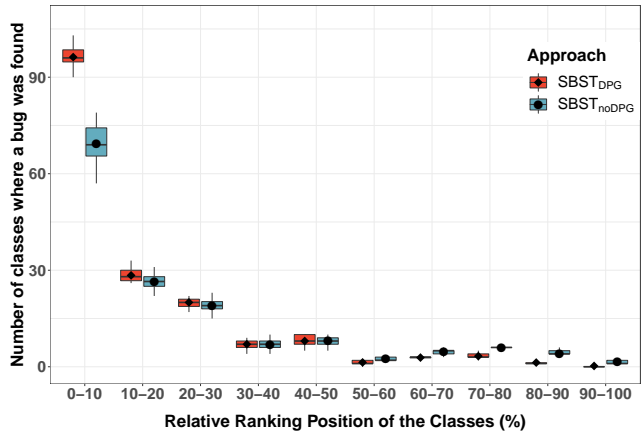}
    \caption{The number of classes where a bug was found by the 2 approaches, grouped by the relative ranking positions (\%) of the classes in the project at T = $15*N$ seconds}
    \label{fig:rq2_results_1}
\end{figure}

\begin{table}[h!]
\caption{Summary of the bug finding results grouped by the relative ranking position (\%) of the classes in the project at T = $15*N$ seconds.}
\label{table:budgetAlloc}
\begin{adjustbox}{max width=\textwidth}
\begin{tabular}{c|r|r|r|r}
\hline
\multirow{3}{*}{\begin{tabular}[c]{@{}c@{}} Rank\\ (\%)\end{tabular}} & \multirow{3}{*}{\begin{tabular}[c]{@{}c@{}}\# Buggy\\ Classes\end{tabular}} & \multirow{3}{*}{\begin{tabular}[c]{@{}c@{}}Avg.\\ Time\\ Budget\end{tabular}} & \multicolumn{2}{c}{\multirow{2}{*}{\begin{tabular}[c]{@{}c@{}} Mean number of classes\\ where a bug was found \end{tabular}}}\\
 &   &   & \multicolumn{2}{c}{}                                                                      \\ \cline{4-5}
 &   &  & SBST$_{DPG}$  & SBST$_{noDPG}$  \\ \hline
0 - 10   & 266   & 66.61   & \textbf{96.25}& 69.45         \\ 
10 - 20 & 72   & 20.76    & \textbf{28.40}   & 26.40   \\ 
20 - 30  & 63 & 16.43  & \textbf{20.00}   & 19.00   \\ 
30 - 40 & 25  & 16.00   & \textbf{7.00}  & 6.75       \\
40 - 50  & 26   & 16.00& \textbf{8.00}    & \textbf{8.00}  \\
50 - 60  & 13    & 2.00  & 1.35                  & \textbf{2.35}                         \\ 
60 - 70    & 16  & 2.00  & 2.85  & \textbf{4.70}    \\ 
70 - 80   & 12 & 2.00 & 3.30  & \textbf{5.95}         \\ 
80 - 90   & 13    & 2.00 & 1.30  & \textbf{4.10}    \\ 
90 - 100 & 5  & 2.00  & 0.25 & \textbf{1.45}   \\ \hline
\end{tabular}
\end{adjustbox}
\end{table}

To further analyse the differences between the two approaches, Figure \ref{fig:rq2_results_1} reports the distribution of the number of classes where a bug was found across 20 runs for the 2 approaches grouped by the relative ranking position produced by Schwa at total time budget of 15 seconds per class. Relative ranking position is the normalised rank of the respective class as described in Algorithm \ref{algo:expowdrank}.

We observe that when the buggy classes are correctly ranked at the top by Schwa, and allocated more time by BADS, the performance of SBST$_{DPG}$ is significantly better than the baseline method. More than half of the buggy classes (52\%) are ranked in the top 10\% of the project by Schwa, as shown in Table~\ref{table:budgetAlloc}, and allocated 66.61 seconds of time budget on average by BADS. Around 36\% of the buggy classes are ranked in the 10-50\% of the projects. BADS employs an exponential function to largely favour a smaller number of highly likely to be defective classes and allocates an adequate amount of time to the moderately defective classes. 

Only 12\% of the buggy classes are ranked below the first half of the project. BADS assumes not all classes in a project are defective and follows the 2-Tier approach to optimise the budget allocation for the project. Thus, all the classes in the \textit{second tier} which contains the classes that are ranked as less likely to be buggy, get a very small time budget (2 seconds). Unsurprisingly, SBST$_{noDPG}$ found more bugs out of these 59 buggy classes than SBST$_{DPG}$. This indicates that the defect predictor's accuracy is key to the better performance of SBST$_{DPG}$ and there is potential to improve our approach further. 

For completeness, we also measure and present the number of true positives, false negatives, and recall of Schwa. Based on the 0.5 threshold, i.e., if the defect score is greater than or equal to 0.5 then the class is buggy and it is non-buggy if the defect score is less than 0.5, Schwa labels 436 buggy classes correctly (true positives) and mislabels 75 buggy classes (false negatives). Hence, Schwa achieves a recall of 0.85.

The defect predictor (i.e., Schwa) and BADS modules add an overhead to SBST$_{DPG}$. While this overhead is accounted in the time budget allocation in SBST$_{DPG}$, we also report the time spent by the defect predictor and BADS modules together. Schwa and BADS spent 0.68 seconds per class on average (standard deviation $=$ 0.4 seconds), which translates to a 4.53\% and 2.27\% overhead in 15 and 30 seconds per class time budgets respectively. Therefore, this shows the overhead introduced by Schwa and BADS in SBST$_{DPG}$ is very small and negligible.

\subsection*{RQ2. Does SBST$_{DPG}$ find more unique bugs?}

\begin{table*}[ht!]
\centering
\caption{Success rate for each method at $15*N$ total time budget. Bug IDs that were found by only one approach are highlighted with different colours;\footnotesize{ \colorbox{cellud}{SBST$_{DPG}$} and \colorbox{cellue}{SBST$_{noDPG}$}.}}
\label{table:rq2_results_1}
\begin{adjustbox}{max width=\textwidth}
\renewcommand{\arraystretch}{1.1}
\begin{tabular}{c|c|c|c|c}
{\pgfplotstabletypeset[
    color cells={min=0,max=1.00},
    col sep=comma,
    /pgf/number format/fixed,
    /pgfplots/colormap={whitegreenyellow}{[1cm] rgb255(0cm)=(252,252,255) rgb255(1cm)=(255,235,132) rgb255(2cm)=(99,190,123)},
    columns/Bug ID/.style={reset styles,string type},
    every head row/.style={before row=\hline,after row=\hline},
    every last row/.style={after row=\hline},
    display columns/0/.style={column type = {l}},
    display columns/1/.style={column type = {|r}},
    display columns/2/.style={column type = {|r}},
]{
Bug ID,SBST$_{DPG}$,SBST$_{noDPG}$
Lang-1,1,0.45
Lang-4,0.9,1
\cellcolor{cellue}Lang-5,0,0.2
Lang-7,1,1
Lang-8,0.1,0.1
Lang-9,0.95,1
Lang-10,0.95,0.8
Lang-11,0.8,0.95
Lang-12,0.2,0.8
\cellcolor{cellud}Lang-14,0.05,0
\cellcolor{cellud}Lang-17,0.05,0
Lang-18,0.5,0.3
Lang-19,0.05,0.7
Lang-20,0.8,0.4
Lang-21,0.1,0.1
Lang-22,0.55,0.8
Lang-23,1,0.95
Lang-27,0.8,0.75
Lang-28,0.05,0.05
Lang-32,1,1
Lang-33,1,1
Lang-34,1,0.9
Lang-35,1,0.3
Lang-36,1,1
Lang-37,0.65,0.2
Lang-39,1,0.95
Lang-41,0.7,1
Lang-44,0.85,0.65
Lang-45,1,1
Lang-46,0.5,1
Lang-47,0.95,0.9
Lang-49,0.55,0.4
Lang-50,0.3,0.3
Lang-51,0.1,0.05
Lang-52,1,1
Lang-53,0.3,0.15
Lang-54,0.05,0.05
\cellcolor{cellud}Lang-55,0.05,0
Lang-57,1,1
\cellcolor{cellue}Lang-58,0,0.05
Lang-59,1,0.95
Lang-60,0.75,0.3
Lang-61,1,0.25
Lang-65,1,0.95
Math-1,1,1
\cellcolor{cellue}Math-2,0,0.1
Math-3,0.55,1
Math-4,1,1
Math-5,0.45,0.95
Math-6,1,1
}}&
{\pgfplotstabletypeset[
    color cells={min=0,max=1.00},
    col sep=comma,
    /pgf/number format/fixed,
    /pgfplots/colormap={whitegreenyellow}{[1cm] rgb255(0cm)=(252,252,255) rgb255(1cm)=(255,235,132) rgb255(2cm)=(99,190,123)},
    columns/Bug ID/.style={reset styles,string type},
    every head row/.style={before row=\hline,after row=\hline},
    every last row/.style={after row=\hline},
    display columns/0/.style={column type = {l}},
    display columns/1/.style={column type = {|r}},
    display columns/2/.style={column type = {|r}},
]{
Bug ID,SBST$_{DPG}$,SBST$_{noDPG}$
Math-9,0.7,0.6
\cellcolor{cellud}Math-10,0.1,0
Math-11,0.95,1
Math-14,1,1
\cellcolor{cellue}Math-16,0,0.05
Math-21,0.05,0.45
Math-22,1,1
Math-23,0.95,0.8
Math-24,0.9,0.85
\cellcolor{cellud}Math-25,0.1,0
Math-26,1,1
Math-27,0.6,0.65
\cellcolor{cellud}Math-28,0.05,0
Math-29,0.9,1
Math-32,1,1
Math-33,0.45,0.35
Math-35,1,1
Math-36,0.2,0.1
Math-37,1,1
Math-40,1,0.95
Math-41,0.25,0.4
Math-42,0.95,0.95
Math-43,0.45,0.55
\cellcolor{cellue}Math-45,0,0.3
Math-46,1,1
Math-47,1,0.95
Math-48,0.65,0.75
Math-49,0.8,0.75
Math-50,0.75,0.3
Math-51,0.35,0.25
Math-52,0.65,0.6
Math-53,1,1
Math-55,1,1
Math-56,1,0.9
Math-59,1,1
Math-60,0.95,0.95
Math-61,1,1
Math-63,1,0.4
\cellcolor{cellud}Math-64,0.05,0
Math-65,0.25,0.25
Math-66,1,1
Math-67,1,1
Math-68,1,1
Math-70,1,1
Math-71,0.6,0.35
Math-72,0.5,0.45
Math-73,0.75,1
Math-75,1,0.9
Math-76,0.15,0.05
Math-77,1,1
}}&
{\pgfplotstabletypeset[
    color cells={min=0,max=1.00},
    col sep=comma,
    /pgf/number format/fixed,
    /pgfplots/colormap={whitegreenyellow}{[1cm] rgb255(0cm)=(252,252,255) rgb255(1cm)=(255,235,132) rgb255(2cm)=(99,190,123)},
    columns/Bug ID/.style={reset styles,string type},
    every head row/.style={before row=\hline,after row=\hline},
    every last row/.style={after row=\hline},
    display columns/0/.style={column type = {l}},
    display columns/1/.style={column type = {|r}},
    display columns/2/.style={column type = {|r}},
]{
Bug ID,SBST$_{DPG}$,SBST$_{noDPG}$
Math-78,0.6,0.6
Math-79,0.15,0.05
\cellcolor{cellud}Math-80,0.3,0
\cellcolor{cellud}Math-81,0.15,0
Math-83,0.9,1
\cellcolor{cellud}Math-84,0.15,0
Math-85,1,1
Math-86,0.95,0.85
Math-87,0.95,1
Math-88,0.75,0.7
Math-89,1,1
Math-90,1,1
Math-92,1,1
Math-93,0.35,0.25
\cellcolor{cellud}Math-94,0.35,0
Math-95,1,1
Math-96,1,1
Math-97,1,1
Math-98,1,0.85
Math-100,1,1
Math-101,0.2,1
Math-102,0.75,0.5
Math-103,1,1
Math-104,0.5,0.4
Math-105,1,1
\cellcolor{cellud}Math-106,0.15,0
Time-1,1,1
Time-2,0.85,1
Time-3,0.15,0.05
\cellcolor{cellue}Time-4,0,0.3
Time-5,1,1
Time-6,1,0.8
\cellcolor{cellud}Time-7,0.15,0
Time-8,1,0.7
Time-9,1,1
Time-10,0.1,0.1
Time-11,1,1
Time-12,1,0.55
Time-13,0.5,0.05
\cellcolor{cellue}Time-14,0,0.95
Time-15,0.4,0.3
\cellcolor{cellud}Time-16,0.15,0
Time-17,1,0.55
\cellcolor{cellue}Time-22,0,0.25
\cellcolor{cellue}Time-23,0,0.2
\cellcolor{cellue}Time-24,0,0.45
Time-26,0.1,0.05
Time-27,0.15,0.5
Chart-1,0.2,0.05
\cellcolor{cellud}Chart-2,0.05,0
}}&
{\pgfplotstabletypeset[
    color cells={min=0,max=1.00},
    col sep=comma,
    /pgf/number format/fixed,
    /pgfplots/colormap={whitegreenyellow}{[1cm] rgb255(0cm)=(252,252,255) rgb255(1cm)=(255,235,132) rgb255(2cm)=(99,190,123)},
    columns/Bug ID/.style={reset styles,string type},
    every head row/.style={before row=\hline,after row=\hline},
    every last row/.style={after row=\hline},
    display columns/0/.style={column type = {l}},
    display columns/1/.style={column type = {|r}},
]{
Bug ID,SBST$_{DPG}$,SBST$_{noDPG}$
Chart-3,0.9,0.15
Chart-4,0.85,0.3
Chart-5,0.35,1
Chart-6,0.8,1
Chart-7,0.3,0.25
Chart-8,1,1
Chart-10,1,1
Chart-11,0.2,1
Chart-12,0.9,0.5
Chart-13,0.9,0.2
Chart-14,1,1
Chart-15,1,0.9
Chart-16,1,1
Chart-17,1,1
Chart-18,1,1
Chart-19,1,0.15
Chart-20,0.5,0.1
Chart-21,0.55,0.05
Chart-22,1,1
Chart-23,1,1
\cellcolor{cellue}Chart-24,0,1
Mockito-2,1,1
Mockito-17,1,1
Mockito-29,0.85,0.95
Mockito-35,1,1
\cellcolor{cellud}Closure-6,0.05,0
Closure-7,0.35,0.1
Closure-9,0.6,0.15
Closure-12,0.3,0.1
\cellcolor{cellue}Closure-19,0,0.1
Closure-21,0.9,0.35
Closure-22,0.5,0.5
Closure-26,0.5,0.4
Closure-27,0.25,0.1
Closure-28,1,1
Closure-30,1,0.95
Closure-33,1,0.5
\cellcolor{cellud}Closure-34,0.05,0
Closure-39,1,0.6
\cellcolor{cellud}Closure-41,0.1,0
\cellcolor{cellud}Closure-43,0.05,0
Closure-46,1,1
\cellcolor{cellud}Closure-48,0.1,0
Closure-49,0.45,0.5
Closure-52,0.4,0.1
Closure-54,1,0.8
Closure-56,0.95,1
\cellcolor{cellud}Closure-60,0.1,0
Closure-65,0.9,0.45
Closure-72,0.2,0.3
}}&
{\pgfplotstabletypeset[
    color cells={min=0,max=1.00},
    col sep=comma,
    /pgf/number format/fixed,
    /pgfplots/colormap={whitegreenyellow}{[1cm] rgb255(0cm)=(252,252,255) rgb255(1cm)=(255,235,132) rgb255(2cm)=(99,190,123)},
    columns/Bug ID/.style={reset styles,string type},
    every head row/.style={before row=\hline,after row=\hline},
    every last row/.style={after row=\hline},
    display columns/0/.style={column type = {l}},
    display columns/1/.style={column type = {|r}},
    display columns/2/.style={column type = {|r}},
]{
Bug ID,SBST$_{DPG}$,SBST$_{noDPG}$
Closure-73,1,1
Closure-77,0.7,0.25
\cellcolor{cellud}Closure-78,0.05,0
Closure-79,1,0.85
\cellcolor{cellud}Closure-80,0.2,0
\cellcolor{cellud}Closure-81,0.35,0
Closure-82,1,1
\cellcolor{cellud}Closure-86,0.15,0
\cellcolor{cellud}Closure-89,0.05,0
\cellcolor{cellud}Closure-91,0.15,0
\cellcolor{cellud}Closure-94,0.25,0
Closure-104,0.95,0.5
Closure-106,1,0.95
Closure-108,0.8,0.2
Closure-110,0.95,1
\cellcolor{cellud}Closure-112,0.1,0
Closure-113,0.25,0.05
\cellcolor{cellue}Closure-114,0,0.1
Closure-115,0.3,0.25
Closure-116,0.2,0.1
Closure-117,0.4,0.05
\cellcolor{cellud}Closure-119,0.25,0
Closure-120,0.2,0.1
Closure-121,0.55,0.2
\cellcolor{cellud}Closure-122,0.05,0
Closure-123,0.15,0.1
\cellcolor{cellud}Closure-125,0.45,0
Closure-128,0.15,0.1
Closure-129,0.2,0.05
Closure-131,0.15,0.9
Closure-137,0.95,1
Closure-139,0.15,0.05
Closure-140,0.85,0.25
\cellcolor{cellud}Closure-141,0.3,0
Closure-144,0.3,0.1
\cellcolor{cellud}Closure-146,0.15,0
Closure-150,0.45,0.1
Closure-151,1,1
Closure-160,0.55,0.05
Closure-164,0.35,0.45
Closure-165,0.95,0.8
\cellcolor{cellud}Closure-167,0.35,0
\cellcolor{cellue}Closure-169,0,0.05
Closure-170,0.2,0.2
Closure-171,0.9,0.05
Closure-172,0.65,0.15
Closure-173,1,0.5
Closure-174,1,1
Closure-175,0.75,0.15
Closure-176,0.1,0.1
}}
\end{tabular}


\end{adjustbox}
\end{table*}

To investigate how our approach performs against each bug, we present an overview of the success rates for each SBST method at total time budget of 15 seconds per class in Table~\ref{table:rq2_results_1}.
Success rate is the ratio of runs where the bug was detected. Due to space limitation, we omit the entries for bugs where none of the approaches were able to find the bug. We also highlight the bugs that were detected by only one approach. As can be seen from Table~\ref{table:rq2_results_1}, our approach outperforms the benchmark in terms of the success rates for most of the bugs. 

\begin{table}[h!]
\renewcommand{\arraystretch}{1.1}
\caption{Summary of the bug finding results at T = $15*N$.}
\label{table:rq2_results_3}
\begin{adjustbox}{max width=0.46\textwidth}
\begin{tabular}{l|c|c|c|c}
\hline
\multirow{3}{*}{} & \multirow{3}{*}{\begin{tabular}[c]{@{}c@{}}Bugs\\ found\end{tabular}} & \multirow{3}{*}{\begin{tabular}[c]{@{}c@{}}Unique \\bugs\end{tabular}} & \multirow{3}{*}{\begin{tabular}[c]{@{}c@{}}Bugs found \\ in every run\end{tabular}} & \multirow{3}{*}{\begin{tabular}[c]{@{}c@{}}Bugs found \\ more often\end{tabular}} \\
&  &    &  &                                                                                   \\
    &     &       &       &                                                            \\ \hline
SBST$_{DPG}$      & \textbf{236}    & \textbf{35}   & \textbf{84}    & \textbf{127}   \\
SBST$_{noDPG}$   & 215    & 14    & 76     & 47 \\ \hline
\end{tabular}
\end{adjustbox}
\end{table}

This observation can be confirmed with the summary of the results which we report in Table \ref{table:rq2_results_3}. What is particularly interesting to observe from the more granular representation of the results in Table~\ref{table:rq2_results_1} is the high number of bugs where our approach has 100\% success rate, which means that SBST$_{DPG}$ finds the respective bugs in all the runs. This is an indication of the robustness of our approach.

Certain bugs are harder to find than others. Out of the 20 runs for each SBST approach, if a bug is only detected by one of the approaches, we call it a unique bug. The reason why we pay special attention to unique bugs is because they are an indication of the ability of the testing technique to discover what cannot be discovered otherwise in the given time budget, which is an important strength \citep{habib2018many}. SBST$_{DPG}$ found 236 bugs altogether, which is 54.38\% of the total bugs, whereas SBST$_{noDPG}$ found only 215 (49.54\%) bugs. SBST$_{DPG}$ found 35 unique bugs that SBST$_{noDPG}$ could not find in any of the runs. On the other hand, SBST$_{noDPG}$ found only 14 such unique bugs. 30 out of these 35 bugs have buggy classes ranked in the top 10\% of the project by Schwa, and the other 5 bugs in 10-50\% of the project. We observe similar results at total time budget of 30 seconds per class as well, where SBST$_{DPG}$ found 32 unique bugs, while SBST$_{noDPG}$ was only able to find 13 unique bugs.

SBST$_{DPG}$ found 127 bugs more times than SBST$_{noDPG}$, while for SBST$_{noDPG}$, this is only 47. 92 out of these 127 bugs have buggy classes ranked in the top 10\% of the project and the other 35 bugs in 10-50\% of the project.

If we consider a bug as found only if all the runs by an approach find the bug (success rate = 1.00), then the number of bugs found by SBST$_{DPG}$ and SBST$_{noDPG}$ become 84 and 76. There are 27 bugs which only SBST$_{DPG}$ detected them in all of the runs. 

\begin{tcolorbox}
In summary, SBST$_{DPG}$ finds 35 more unique bugs compared to the benchmark approach. Furthermore, it finds a large number of bugs more frequently than the baseline. Thus, this suggests that the superior performance of SBST$_{DPG}$ is supported by both its capability of finding new bugs which are not exposed by the baseline and the robustness of the approach.
\end{tcolorbox}

We pick Math-94 and Time-8 bugs and investigate the tests generated by the 2 approaches. Figure \ref{listing:math-94-buggy-code} shows the buggy code snippet of \code{MathUtils} class from Math-94. The \code{if} condition at line 412 is placed to check if either \code{u} or \code{v} is zero. This is a classic example of a bug due to an integer overflow. Assume the method is called with the following inputs \code{MathUtils.$gcd$(1073741824, 1032)}. Then, the \code{if} condition at line 412 is expected to be evaluated to \code{false} since both \code{u(1073741824)} and \code{v(1032)} are non-zeros. However, the multiplication of \code{u} and \code{v} causes an integer overflow to zero, and the \code{if} condition at line 412 is evaluated to \code{true}. Figure \ref{listing:math-94-fixed-code} shows the same code snippet of \code{MathUtils} class after the patch is applied. To detect this bug, a test should not only cover the \code{true} branch of the \code{if} condition at line 412, but also pass the non-zero arguments \code{u} and \code{v} such that their multiplication causes an integer overflow to zero. 

The fitness function for the \code{true} branch of the \code{if} condition at line 412 is $u*v/(u*v + 1)$, and it tends to reward the test inputs \code{u} and \code{v} whose multiplication is closer to zero more than the ones whose multiplication is closer to causing an integer overflow to zero. For an example, suppose we have two \textit{individuals} $u=2,v=3$ and $u=12085,v=1241$ in the current generation. The fitness of the first and second \textit{individuals} will be $6/(6+1)$ and $14997485/(14997485+1)$. Thus, the first \textit{individual} is considered fitter when compared with the second one, while the second one is closer to detect the bug than the first one. Therefore in a situation like this, we can increase the chances of detecting the bug by allowing the search method to extensively explore the search space of possible test inputs and generate more than one test case (test inputs) for such branches.

SBST$_{noDPG}$ generated 30.75 test cases on average that cover the \code{true} branch of the \code{if} condition at line 412, yet it was not able to detect the bug in any of the runs. Schwa ranked Math-94 bug in the top 10\% of the project and BADS allocated 37 seconds time budget to the search. Then, SBST$_{DPG}$ generated 49.8 test cases on average that cover the said branch. As a result, it was able to find the bug in 7 runs out of 20. Allocating a higher time budget increases the likelihood of detecting the bug since it allows the search method to explore the search space extensively to find the test inputs that can detect the bug.

\begin{figure}[h]
\begin{subfigure}{0.20\textwidth}
\begin{lstlisting}[language=java, firstline=411, lastline=416, firstnumber=411, xleftmargin=0em, columns=fullflexible]
public static int gcd(int u, int v) {
	if (u * v == 0) {
		return (Math.abs(u) + Math.abs(v));
	}
	...
}
\end{lstlisting}
\caption{Buggy code}
\label{listing:math-94-buggy-code}
\end{subfigure}\hspace{0.03\textwidth}
\begin{subfigure}{0.20\textwidth}
\begin{lstlisting}[language=java, firstline=411, lastline=416, firstnumber=411, xrightmargin=-2em, columns=fullflexible]
public static int gcd(int u, int v) {
	if ((u == 0) || (v == 0)) {
		return (Math.abs(u) + Math.abs(v));
	}
	...
}
\end{lstlisting}
\caption{Fixed code}
\label{listing:math-94-fixed-code}
\end{subfigure}
\caption{MathUtils class from Math-94}
\end{figure}

Figure \ref{listing:time-8-buggy-code} shows the buggy code snippet of \code{DateTimeZone} class from Time-8. The \code{forOffsetHoursMinutes} method takes two integer inputs \code{hoursOffset} and \code{minutesOffset}, and returns the \code{DateTimeZone} object for the offset specified by the two inputs. If the method \code{forOffsetHoursMinutes} is called with the inputs \code{hoursOffset=0} and \code{minutesOffset=-15}, then it is expected to return a \code{DateTimeZone} object for the offset $-00:15$. However, the \code{if} condition at line 279 is evaluated to \code{true} and the method throws an \code{IllegalArgumentException} instead. Figure \ref{listing:time-8-fixed-code} shows the same code snippet after the patch is applied. To detect this bug, a test case has to execute the \code{if} conditions at lines 273 and 276 to \code{false}; that is \code{hoursOffset} $\neq 0$ or \code{minutesOffset} $\neq 0$ and \code{hoursOffset} $\in [-23,23]$, and then it has to execute the \code{if} condition at line 279 to \code{true} with a \code{minutesOffset} $\in [-59,-1]$. Moreover, there is a new condition introduced at line 282 in the fixed code to check if the \code{hoursOffset} is positive when the \code{minutesOffset} is negative (see Figure \ref{listing:time-8-fixed-code}). Thus, this adds another constraint to the possible test inputs that can detect the bug, which is \code{hoursOffset} $\leq 0$. Therefore, it is evident that it is hard not only to find the right test inputs to detect the bug, but also to find test inputs to at least cover the buggy code.

As it was the case in Math-94, just covering the buggy code (\code{true} branch of the \code{if} condition at line 279) is not sufficient to detect the Time-8 bug. For an example, test inputs \code{hoursOffset=-4} and \code{minutesOffset=-150} cover the buggy code, however they cannot detect the bug. Therefore, the search method needs more resources to generate more test cases that cover the buggy code such that it eventually finds the right test cases that can detect the bug.
\begin{figure}[h]

\begin{subfigure}{0.48\textwidth}
\begin{lstlisting}[language=java, firstline=272, lastline=284, firstnumber=272]
public static DateTimeZone forOffsetHoursMinutes(int hoursOffset, int minutesOffset) throws IllegalArgumentException {
	if (hoursOffset == 0 && minutesOffset == 0) {
		return DateTimeZone.UTC;
	}
	if (hoursOffset < -23 || hoursOffset > 23) {
		throw new IllegalArgumentException("Hours out of range: " + hoursOffset);
	}
	if (minutesOffset < 0 || minutesOffset > 59) {
		throw new IllegalArgumentException("Minutes out of range: " + minutesOffset);
	}
	int offset = 0;
	...
}
\end{lstlisting}
\caption{Buggy code}
\label{listing:time-8-buggy-code}
\end{subfigure}
\begin{subfigure}{0.48\textwidth}
\begin{lstlisting}[language=java, firstline=272, lastline=287, firstnumber=272]
public static DateTimeZone forOffsetHoursMinutes(int hoursOffset, int minutesOffset) throws IllegalArgumentException {
	if (hoursOffset == 0 && minutesOffset == 0) {
		return DateTimeZone.UTC;
	}
	if (hoursOffset < -23 || hoursOffset > 23) {
		throw new IllegalArgumentException("Hours out of range: " + hoursOffset);
	}
	if (minutesOffset < -59 || minutesOffset > 59) {
		throw new IllegalArgumentException("Minutes out of range: " + minutesOffset);
	}
	if (hoursOffset > 0 && minutesOffset < 0) {
		throw new IllegalArgumentException("Positive hours must not have negative minutes: " + minutesOffset);
	}
	int offset = 0;
	...
}
\end{lstlisting}
\caption{Fixed code}
\label{listing:time-8-fixed-code}
\end{subfigure}
\caption{DateTimeZone class from Time-8}
\end{figure}

Our investigation into the tests generated by the 2 approaches shows that the baseline, SBST$_{noDPG}$, covered the buggy code in 90\% of the runs. SBST$_{noDPG}$ generated 25.78 test cases on average that cover the buggy code and it was able to detect the bug in 14 runs out of 20. Whereas, SBST$_{DPG}$ allocated 75 seconds time budget to the search as Schwa ranked the bug in the top 10\% of the project and generated 109.8 test cases on average that cover the buggy code. As a result, it was able to detect the bug in all of the runs (success rate = 1.00). Therefore, this again confirms the importance of focusing the search more into the buggy classes to increase the likelihood of detecting the bug.

\section{Threats to Validity}

\textbf{Internal Validity.} As outlined in Section~\ref{sec:param_settings_evosuite}, we configure $Dyna\-MOSA$ to generate more than one test case for each target in the SUT, retain all these test cases and disable test suite minimisation. By doing this, we expect to compromise the test suite size in order to maximise the bug detection of SBST. To investigate the benefit of configuring $DynaMOSA$ in this way, we also run the same set of experiments using $DynaMOSA$ with test suite minimisation and equal budget allocation, SBST$_{O}$. We compare its performance against SBST$_{noDPG}$. SBST$_{O}$ finds 85.75 and 93.45 bugs on average at total time budget of $15$ and $30$ seconds per class. SBST$_{noDPG}$ outperforms SBST$_O$ with an average improvement of 48.2 (+56.2\%) and 73.45 (+78.6\%) more bugs in each case, which are statistically significant according to the Mann-Whitney U-Test (p-value < 0.0001) with a large effect size ($\widehat{A}_{12}$ = 1.00). However, this huge improvement comes with a price, i.e., SBST$_{noDPG}$ produces large test suites. This can be problematic if the developers have to insert the test oracles manually to the generated tests. Thus, we identify this as a potential threat to internal validity and future works need to be done on adapting appropriate test suite minimisation techniques to SBST$_{DPG}$.

To encounter the randomised nature of GA used in $DynaMOSA$, we run the experiments for 20 times and carry out sound statistical tests; two-tailed non-parametric Mann-Whitney U-Test \citep{arcuri2014hitchhiker} and Vargha and Delaney's $\widehat{A}_{12}$ statistic \citep{vargha2000critique}.

The parameter configurations for Schwa and BADS are either the default values or based on the results of the pilot runs. We believe the performance of SBST$_{DPG}$ can be further improved by fine-tuning the parameters of Schwa and BADS. 

We employ an exponential function to allocate time budgets for classes based on the defect scores. As opposed to an exponential allocation, a direct mapping (i.e., linear budget allocation) would have been simple and straight-forward. However, as described in Section~\ref{subsec:exp_time_BADS}, there are only a few number of classes which are actually buggy (i.e., highly likely to be defective) and they need to be allocated more time budget to maximise the bug detection of the test generation tool. Thus, we believe a linear allocation approach is not able to largely favour these small number of classes like the exponential allocation approach does.

\textbf{External Validity.} We use 434 real bugs from Defects4J dataset that are drawn from 6 open source projects. These projects may not represent all program characteristics; especially in industrial projects. Although, Defects4J has been widely used in the literature \citep{shamshiri2015automatically,paterson2019empirical,le2016history,pearson2017evaluating} as a benchmark. Future work needs to be done on applying SBST$_{DPG}$ on other bugs datasets. 

EvoSuite generates JUnit test suites for Java programs. Thus, we may not be able to generalise the conclusions to other programming languages. However, the concept we introduced in this research is not language dependent and can be applied to other programming languages as well.

\section{Conclusion}

We introduce defect prediction guided SBST (SBST$_{DPG}$) that combines class level defect prediction and Search-Based Software Testing to efficiently find bugs in a resource constrained environment. SBST$_{DPG}$ employs a budget allocation algorithm, Budget Allocation Based on Defect Scores (BADS), to allocate time budgets for classes based on their likelihood of defectiveness. We validate our approach against 434 real bugs from Defects4J dataset. Our empirical evaluation demonstrates that in a resource constrained environment, when given a tight time budget, SBST$_{DPG}$ is significantly more efficient than the state of the art approach with a large effect size. In particular, SBST$_{DPG}$ finds 13.1\% more bugs on average compared to the state of the art SBST approach when they are given a tight time budget of 15 seconds per class. Further analysis of the results finds that the superior performance of SBST$_{DPG}$ is supported by its ability to find more unique bugs which otherwise remain undetected.

We aim to extend our work in the following directions as future work; i) employ a defect predictor which uses different features to produce predictions, ii) adapt an appropriate test suite minimisation strategy to address the generation of larger test suites, and iii) validate SBST$_{DPG}$ against other bugs datasets.

\section*{Acknowledgements}
This work was partially funded by the Australian Research Council (ARC) through a Discovery Early Career Researcher Award (DE190100046).

\clearpage
\bibliographystyle{ACM-Reference-Format}
\bibliography{reference}


\begin{thebibliography}{61}


\ifx \showCODEN    \undefined \def \showCODEN     #1{\unskip}     \fi
\ifx \showDOI      \undefined \def \showDOI       #1{#1}\fi
\ifx \showISBNx    \undefined \def \showISBNx     #1{\unskip}     \fi
\ifx \showISBNxiii \undefined \def \showISBNxiii  #1{\unskip}     \fi
\ifx \showISSN     \undefined \def \showISSN      #1{\unskip}     \fi
\ifx \showLCCN     \undefined \def \showLCCN      #1{\unskip}     \fi
\ifx \shownote     \undefined \def \shownote      #1{#1}          \fi
\ifx \showarticletitle \undefined \def \showarticletitle #1{#1}   \fi
\ifx \showURL      \undefined \def \showURL       {\relax}        \fi
\providecommand\bibfield[2]{#2}
\providecommand\bibinfo[2]{#2}
\providecommand\natexlab[1]{#1}
\providecommand\showeprint[2][]{arXiv:#2}

\bibitem[\protect\citeauthoryear{Aleti and Grunske}{Aleti and Grunske}{2015}]%
        {aleti2015test}
\bibfield{author}{\bibinfo{person}{Aldeida Aleti} {and} \bibinfo{person}{Lars
  Grunske}.} \bibinfo{year}{2015}\natexlab{}.
\newblock \showarticletitle{Test data generation with a Kalman filter-based
  adaptive genetic algorithm}.
\newblock \bibinfo{journal}{\emph{Journal of Systems and Software}}
  \bibinfo{volume}{103} (\bibinfo{year}{2015}), \bibinfo{pages}{343--352}.
\newblock


\bibitem[\protect\citeauthoryear{Aleti, Moser, and Grunske}{Aleti
  et~al\mbox{.}}{2017}]%
        {aleti2017analysing}
\bibfield{author}{\bibinfo{person}{Aldeida Aleti}, \bibinfo{person}{Irene
  Moser}, {and} \bibinfo{person}{Lars Grunske}.}
  \bibinfo{year}{2017}\natexlab{}.
\newblock \showarticletitle{Analysing the fitness landscape of search-based
  software testing problems}.
\newblock \bibinfo{journal}{\emph{Automated Software Engineering}}
  \bibinfo{volume}{24}, \bibinfo{number}{3} (\bibinfo{year}{2017}),
  \bibinfo{pages}{603--621}.
\newblock


\bibitem[\protect\citeauthoryear{Almasi, Hemmati, Fraser, Arcuri, and
  Benefelds}{Almasi et~al\mbox{.}}{2017}]%
        {almasi2017industrial}
\bibfield{author}{\bibinfo{person}{M~Moein Almasi}, \bibinfo{person}{Hadi
  Hemmati}, \bibinfo{person}{Gordon Fraser}, \bibinfo{person}{Andrea Arcuri},
  {and} \bibinfo{person}{J{\=a}nis Benefelds}.}
  \bibinfo{year}{2017}\natexlab{}.
\newblock \showarticletitle{An industrial evaluation of unit test generation:
  Finding real faults in a financial application}. In
  \bibinfo{booktitle}{\emph{Proceedings of the 39th International Conference on
  Software Engineering: Software Engineering in Practice Track}}. IEEE Press,
  \bibinfo{pages}{263--272}.
\newblock


\bibitem[\protect\citeauthoryear{Alshahwan, Gao, Harman, Jia, Mao, Mols, Tei,
  and Zorin}{Alshahwan et~al\mbox{.}}{2018}]%
        {alshahwan2018deploying}
\bibfield{author}{\bibinfo{person}{Nadia Alshahwan}, \bibinfo{person}{Xinbo
  Gao}, \bibinfo{person}{Mark Harman}, \bibinfo{person}{Yue Jia},
  \bibinfo{person}{Ke Mao}, \bibinfo{person}{Alexander Mols},
  \bibinfo{person}{Taijin Tei}, {and} \bibinfo{person}{Ilya Zorin}.}
  \bibinfo{year}{2018}\natexlab{}.
\newblock \showarticletitle{Deploying search based software engineering with
  Sapienz at Facebook}. In \bibinfo{booktitle}{\emph{International Symposium on
  Search Based Software Engineering}}. Springer, \bibinfo{pages}{3--45}.
\newblock


\bibitem[\protect\citeauthoryear{Arcuri and Briand}{Arcuri and Briand}{2014}]%
        {arcuri2014hitchhiker}
\bibfield{author}{\bibinfo{person}{Andrea Arcuri} {and} \bibinfo{person}{Lionel
  Briand}.} \bibinfo{year}{2014}\natexlab{}.
\newblock \showarticletitle{A Hitchhiker's guide to statistical tests for
  assessing randomized algorithms in software engineering}.
\newblock \bibinfo{journal}{\emph{Software Testing, Verification and
  Reliability}} \bibinfo{volume}{24}, \bibinfo{number}{3}
  (\bibinfo{year}{2014}), \bibinfo{pages}{219--250}.
\newblock


\bibitem[\protect\citeauthoryear{Arcuri, Campos, and Fraser}{Arcuri
  et~al\mbox{.}}{2016}]%
        {arcuri2016unit}
\bibfield{author}{\bibinfo{person}{Andrea Arcuri}, \bibinfo{person}{Jos{\'e}
  Campos}, {and} \bibinfo{person}{Gordon Fraser}.}
  \bibinfo{year}{2016}\natexlab{}.
\newblock \showarticletitle{Unit test generation during software development:
  Evosuite plugins for maven, intellij and jenkins}. In
  \bibinfo{booktitle}{\emph{2016 IEEE International Conference on Software
  Testing, Verification and Validation (ICST)}}. IEEE,
  \bibinfo{pages}{401--408}.
\newblock


\bibitem[\protect\citeauthoryear{Arcuri and Fraser}{Arcuri and Fraser}{2013}]%
        {arcuri2013parameter}
\bibfield{author}{\bibinfo{person}{Andrea Arcuri} {and} \bibinfo{person}{Gordon
  Fraser}.} \bibinfo{year}{2013}\natexlab{}.
\newblock \showarticletitle{Parameter tuning or default values? An empirical
  investigation in search-based software engineering}.
\newblock \bibinfo{journal}{\emph{Empirical Software Engineering}}
  \bibinfo{volume}{18}, \bibinfo{number}{3} (\bibinfo{year}{2013}),
  \bibinfo{pages}{594--623}.
\newblock


\bibitem[\protect\citeauthoryear{Basili, Briand, and Melo}{Basili
  et~al\mbox{.}}{1996}]%
        {basili1996validation}
\bibfield{author}{\bibinfo{person}{Victor~R Basili}, \bibinfo{person}{Lionel~C.
  Briand}, {and} \bibinfo{person}{Walc{\'e}lio~L Melo}.}
  \bibinfo{year}{1996}\natexlab{}.
\newblock \showarticletitle{A validation of object-oriented design metrics as
  quality indicators}.
\newblock \bibinfo{journal}{\emph{IEEE Transactions on software engineering}}
  \bibinfo{volume}{22}, \bibinfo{number}{10} (\bibinfo{year}{1996}),
  \bibinfo{pages}{751--761}.
\newblock


\bibitem[\protect\citeauthoryear{Broy, Kruger, Pretschner, and Salzmann}{Broy
  et~al\mbox{.}}{2007}]%
        {broy2007engineering}
\bibfield{author}{\bibinfo{person}{Manfred Broy}, \bibinfo{person}{Ingolf~H
  Kruger}, \bibinfo{person}{Alexander Pretschner}, {and}
  \bibinfo{person}{Christian Salzmann}.} \bibinfo{year}{2007}\natexlab{}.
\newblock \showarticletitle{Engineering automotive software}.
\newblock \bibinfo{journal}{\emph{Proc. IEEE}} \bibinfo{volume}{95},
  \bibinfo{number}{2} (\bibinfo{year}{2007}), \bibinfo{pages}{356--373}.
\newblock


\bibitem[\protect\citeauthoryear{Caglayan, Turhan, Bener, Habayeb, Miransky,
  and Cialini}{Caglayan et~al\mbox{.}}{2015}]%
        {caglayan2015merits}
\bibfield{author}{\bibinfo{person}{Bora Caglayan}, \bibinfo{person}{Burak
  Turhan}, \bibinfo{person}{Ayse Bener}, \bibinfo{person}{Mayy Habayeb},
  \bibinfo{person}{Andriy Miransky}, {and} \bibinfo{person}{Enzo Cialini}.}
  \bibinfo{year}{2015}\natexlab{}.
\newblock \showarticletitle{Merits of organizational metrics in defect
  prediction: an industrial replication}. In
  \bibinfo{booktitle}{\emph{Proceedings of the 37th International Conference on
  Software Engineering-Volume 2}}. IEEE Press, \bibinfo{pages}{89--98}.
\newblock


\bibitem[\protect\citeauthoryear{Campos, Arcuri, Fraser, and Abreu}{Campos
  et~al\mbox{.}}{2014}]%
        {campos2014continuous}
\bibfield{author}{\bibinfo{person}{Jos{\'e} Campos}, \bibinfo{person}{Andrea
  Arcuri}, \bibinfo{person}{Gordon Fraser}, {and} \bibinfo{person}{Rui Abreu}.}
  \bibinfo{year}{2014}\natexlab{}.
\newblock \showarticletitle{Continuous test generation: enhancing continuous
  integration with automated test generation}. In
  \bibinfo{booktitle}{\emph{Proceedings of the 29th ACM/IEEE international
  conference on Automated software engineering}}. ACM, \bibinfo{pages}{55--66}.
\newblock


\bibitem[\protect\citeauthoryear{Campos, Panichella, and Fraser}{Campos
  et~al\mbox{.}}{2019}]%
        {campos2019evosuite}
\bibfield{author}{\bibinfo{person}{Jos{\'e} Campos}, \bibinfo{person}{Annibale
  Panichella}, {and} \bibinfo{person}{Gordon Fraser}.}
  \bibinfo{year}{2019}\natexlab{}.
\newblock \showarticletitle{EvoSuiTE at the SBST 2019 tool competition}. In
  \bibinfo{booktitle}{\emph{Proceedings of the 12th International Workshop on
  Search-Based Software Testing}}. IEEE Press, \bibinfo{pages}{29--32}.
\newblock


\bibitem[\protect\citeauthoryear{Dam, Pham, Ng, Tran, Grundy, Ghose, Kim, and
  Kim}{Dam et~al\mbox{.}}{2019}]%
        {dam2019lessons}
\bibfield{author}{\bibinfo{person}{Hoa~Khanh Dam}, \bibinfo{person}{Trang
  Pham}, \bibinfo{person}{Shien~Wee Ng}, \bibinfo{person}{Truyen Tran},
  \bibinfo{person}{John Grundy}, \bibinfo{person}{Aditya Ghose},
  \bibinfo{person}{Taeksu Kim}, {and} \bibinfo{person}{Chul-Joo Kim}.}
  \bibinfo{year}{2019}\natexlab{}.
\newblock \showarticletitle{Lessons learned from using a deep tree-based model
  for software defect prediction in practice}. In
  \bibinfo{booktitle}{\emph{Proceedings of the 16th International Conference on
  Mining Software Repositories}}. IEEE Press, \bibinfo{pages}{46--57}.
\newblock


\bibitem[\protect\citeauthoryear{de~Freitas}{de~Freitas}{2015}]%
        {de2015software}
\bibfield{author}{\bibinfo{person}{Paulo Andr{\'e}~Faria de Freitas}.}
  \bibinfo{year}{2015}\natexlab{}.
\newblock \showarticletitle{Software Repository Mining Analytics to Estimate
  Software Component Reliability}.
\newblock  (\bibinfo{year}{2015}).
\newblock


\bibitem[\protect\citeauthoryear{EvoSuite}{EvoSuite}{2019}]%
        {evosuitegithub}
\bibfield{author}{\bibinfo{person}{EvoSuite}.} \bibinfo{year}{2019}\natexlab{}.
\newblock \bibinfo{title}{EvoSuite - automated generation of JUnit test suites
  for Java classes}.
\newblock
\newblock
\urldef\tempurl%
\url{https://github.com/EvoSuite/evosuite}
\showURL{%
\tempurl}
\newblock
\shownote{Last accessed on: 29/11/2019.}


\bibitem[\protect\citeauthoryear{Foundation}{Foundation}{2019}]%
        {commons-math}
\bibfield{author}{\bibinfo{person}{The Apache~Software Foundation}.}
  \bibinfo{year}{2019}\natexlab{}.
\newblock \bibinfo{title}{Apache Commons Math}.
\newblock
\newblock
\urldef\tempurl%
\url{https://github.com/apache/commons-math}
\showURL{%
\tempurl}
\newblock
\shownote{Last accessed on: 19/09/2019.}


\bibitem[\protect\citeauthoryear{Fowler and Foemmel}{Fowler and
  Foemmel}{2006}]%
        {fowler2006continuous}
\bibfield{author}{\bibinfo{person}{Martin Fowler} {and}
  \bibinfo{person}{Matthew Foemmel}.} \bibinfo{year}{2006}\natexlab{}.
\newblock \bibinfo{title}{Continuous integration}.
\newblock
\newblock


\bibitem[\protect\citeauthoryear{Fraser}{Fraser}{2018}]%
        {evosuiteweb}
\bibfield{author}{\bibinfo{person}{Gordon Fraser}.}
  \bibinfo{year}{2018}\natexlab{}.
\newblock \bibinfo{title}{EvoSuite - Automatic Test Suite Generation for Java}.
\newblock
\newblock
\urldef\tempurl%
\url{http://www.evosuite.org/}
\showURL{%
\tempurl}
\newblock
\shownote{Last accessed on: 19/09/2019.}


\bibitem[\protect\citeauthoryear{Fraser and Arcuri}{Fraser and Arcuri}{2011}]%
        {fraser2011evolutionary}
\bibfield{author}{\bibinfo{person}{Gordon Fraser} {and} \bibinfo{person}{Andrea
  Arcuri}.} \bibinfo{year}{2011}\natexlab{}.
\newblock \showarticletitle{Evolutionary generation of whole test suites}. In
  \bibinfo{booktitle}{\emph{2011 11th International Conference on Quality
  Software}}. IEEE, \bibinfo{pages}{31--40}.
\newblock


\bibitem[\protect\citeauthoryear{Fraser and Arcuri}{Fraser and Arcuri}{2012}]%
        {fraser2012whole}
\bibfield{author}{\bibinfo{person}{Gordon Fraser} {and} \bibinfo{person}{Andrea
  Arcuri}.} \bibinfo{year}{2012}\natexlab{}.
\newblock \showarticletitle{Whole test suite generation}.
\newblock \bibinfo{journal}{\emph{IEEE Transactions on Software Engineering}}
  \bibinfo{volume}{39}, \bibinfo{number}{2} (\bibinfo{year}{2012}),
  \bibinfo{pages}{276--291}.
\newblock


\bibitem[\protect\citeauthoryear{{Fraser} and {Arcuri}}{{Fraser} and
  {Arcuri}}{2013}]%
        {fraser2013SBSTtool}
\bibfield{author}{\bibinfo{person}{G. {Fraser}} {and} \bibinfo{person}{A.
  {Arcuri}}.} \bibinfo{year}{2013}\natexlab{}.
\newblock \showarticletitle{EvoSuite at the SBST 2013 Tool Competition}. In
  \bibinfo{booktitle}{\emph{2013 IEEE Sixth International Conference on
  Software Testing, Verification and Validation Workshops}}.
  \bibinfo{pages}{406--409}.
\newblock
\showISSN{null}
\urldef\tempurl%
\url{https://doi.org/10.1109/ICSTW.2013.53}
\showDOI{\tempurl}


\bibitem[\protect\citeauthoryear{Fraser and Arcuri}{Fraser and Arcuri}{2014a}]%
        {fraser2014SBSTtool}
\bibfield{author}{\bibinfo{person}{Gordon Fraser} {and} \bibinfo{person}{Andrea
  Arcuri}.} \bibinfo{year}{2014}\natexlab{a}.
\newblock \showarticletitle{EvoSuite at the Second Unit Testing Tool
  Competition}. In \bibinfo{booktitle}{\emph{Future Internet Testing}},
  \bibfield{editor}{\bibinfo{person}{Tanja~E.J. Vos}, \bibinfo{person}{Kiran
  Lakhotia}, {and} \bibinfo{person}{Sebastian Bauersfeld}} (Eds.).
  \bibinfo{publisher}{Springer International Publishing},
  \bibinfo{address}{Cham}, \bibinfo{pages}{95--100}.
\newblock
\showISBNx{978-3-319-07785-7}


\bibitem[\protect\citeauthoryear{Fraser and Arcuri}{Fraser and Arcuri}{2014b}]%
        {fraser2014large}
\bibfield{author}{\bibinfo{person}{Gordon Fraser} {and} \bibinfo{person}{Andrea
  Arcuri}.} \bibinfo{year}{2014}\natexlab{b}.
\newblock \showarticletitle{A large-scale evaluation of automated unit test
  generation using evosuite}.
\newblock \bibinfo{journal}{\emph{ACM Transactions on Software Engineering and
  Methodology (TOSEM)}} \bibinfo{volume}{24}, \bibinfo{number}{2}
  (\bibinfo{year}{2014}), \bibinfo{pages}{8}.
\newblock


\bibitem[\protect\citeauthoryear{Fraser and Arcuri}{Fraser and Arcuri}{2015}]%
        {fraser20151600}
\bibfield{author}{\bibinfo{person}{Gordon Fraser} {and} \bibinfo{person}{Andrea
  Arcuri}.} \bibinfo{year}{2015}\natexlab{}.
\newblock \showarticletitle{1600 faults in 100 projects: automatically finding
  faults while achieving high coverage with evosuite}.
\newblock \bibinfo{journal}{\emph{Empirical Software Engineering}}
  \bibinfo{volume}{20}, \bibinfo{number}{3} (\bibinfo{year}{2015}),
  \bibinfo{pages}{611--639}.
\newblock


\bibitem[\protect\citeauthoryear{Fraser and Arcuri}{Fraser and Arcuri}{2016}]%
        {fraser2016evosuite}
\bibfield{author}{\bibinfo{person}{Gordon Fraser} {and} \bibinfo{person}{Andrea
  Arcuri}.} \bibinfo{year}{2016}\natexlab{}.
\newblock \showarticletitle{EvoSuite at the SBST 2016 tool competition}. In
  \bibinfo{booktitle}{\emph{2016 IEEE/ACM 9th International Workshop on
  Search-Based Software Testing (SBST)}}. IEEE, \bibinfo{pages}{33--36}.
\newblock


\bibitem[\protect\citeauthoryear{Fraser, Rojas, and Arcuri}{Fraser
  et~al\mbox{.}}{2018}]%
        {fraser2018sbsttool}
\bibfield{author}{\bibinfo{person}{Gordon Fraser},
  \bibinfo{person}{Jos{\'e}~Miguel Rojas}, {and} \bibinfo{person}{Andrea
  Arcuri}.} \bibinfo{year}{2018}\natexlab{}.
\newblock \showarticletitle{Evosuite at the SBST 2018 Tool Competition}. In
  \bibinfo{booktitle}{\emph{Proceedings of the 11th International Workshop on
  Search-Based Software Testing}} (Gothenburg, Sweden)
  \emph{(\bibinfo{series}{SBST '18})}. \bibinfo{publisher}{ACM},
  \bibinfo{address}{New York, NY, USA}, \bibinfo{pages}{34--37}.
\newblock
\showISBNx{978-1-4503-5741-8}
\urldef\tempurl%
\url{https://doi.org/10.1145/3194718.3194729}
\showDOI{\tempurl}


\bibitem[\protect\citeauthoryear{Fraser, Rojas, Campos, and Arcuri}{Fraser
  et~al\mbox{.}}{2017}]%
        {fraser2017sbsttool}
\bibfield{author}{\bibinfo{person}{Gordon Fraser},
  \bibinfo{person}{Jos{\'e}~Miguel Rojas}, \bibinfo{person}{Jos{\'e} Campos},
  {and} \bibinfo{person}{Andrea Arcuri}.} \bibinfo{year}{2017}\natexlab{}.
\newblock \showarticletitle{EvoSuite at the SBST 2017 Tool Competition}. In
  \bibinfo{booktitle}{\emph{Proceedings of the 10th International Workshop on
  Search-Based Software Testing}} (Buenos Aires, Argentina)
  \emph{(\bibinfo{series}{SBST '17})}. \bibinfo{publisher}{IEEE Press},
  \bibinfo{address}{Piscataway, NJ, USA}, \bibinfo{pages}{39--41}.
\newblock
\showISBNx{978-1-5386-2789-1}
\urldef\tempurl%
\url{https://doi.org/10.1109/SBST.2017..6}
\showDOI{\tempurl}


\bibitem[\protect\citeauthoryear{Fraser, Staats, McMinn, Arcuri, and
  Padberg}{Fraser et~al\mbox{.}}{2013}]%
        {fraser2013does}
\bibfield{author}{\bibinfo{person}{Gordon Fraser}, \bibinfo{person}{Matt
  Staats}, \bibinfo{person}{Phil McMinn}, \bibinfo{person}{Andrea Arcuri},
  {and} \bibinfo{person}{Frank Padberg}.} \bibinfo{year}{2013}\natexlab{}.
\newblock \showarticletitle{Does automated white-box test generation really
  help software testers?}. In \bibinfo{booktitle}{\emph{Proceedings of the 2013
  International Symposium on Software Testing and Analysis}}. ACM,
  \bibinfo{pages}{291--301}.
\newblock


\bibitem[\protect\citeauthoryear{Freitas}{Freitas}{2015a}]%
        {schwapypi}
\bibfield{author}{\bibinfo{person}{Andre Freitas}.}
  \bibinfo{year}{2015}\natexlab{a}.
\newblock \bibinfo{title}{Schwa}.
\newblock
\newblock
\urldef\tempurl%
\url{https://pypi.org/project/Schwa}
\showURL{%
\tempurl}
\newblock
\shownote{Last accessed on 16/09/2019.}


\bibitem[\protect\citeauthoryear{Freitas}{Freitas}{2015b}]%
        {schwagithub}
\bibfield{author}{\bibinfo{person}{André Freitas}.}
  \bibinfo{year}{2015}\natexlab{b}.
\newblock \bibinfo{title}{schwa}.
\newblock
\newblock
\urldef\tempurl%
\url{https://github.com/andrefreitas/schwa}
\showURL{%
\tempurl}
\newblock
\shownote{Last accessed on 16/09/2019.}


\bibitem[\protect\citeauthoryear{Gay}{Gay}{2017}]%
        {gay2017generating}
\bibfield{author}{\bibinfo{person}{Gregory Gay}.}
  \bibinfo{year}{2017}\natexlab{}.
\newblock \showarticletitle{Generating effective test suites by combining
  coverage criteria}. In \bibinfo{booktitle}{\emph{International Symposium on
  Search Based Software Engineering}}. Springer, \bibinfo{pages}{65--82}.
\newblock


\bibitem[\protect\citeauthoryear{Giger, D'Ambros, Pinzger, and Gall}{Giger
  et~al\mbox{.}}{2012}]%
        {giger2012method}
\bibfield{author}{\bibinfo{person}{Emanuel Giger}, \bibinfo{person}{Marco
  D'Ambros}, \bibinfo{person}{Martin Pinzger}, {and} \bibinfo{person}{Harald~C
  Gall}.} \bibinfo{year}{2012}\natexlab{}.
\newblock \showarticletitle{Method-level bug prediction}. In
  \bibinfo{booktitle}{\emph{Proceedings of the 2012 ACM-IEEE International
  Symposium on Empirical Software Engineering and Measurement}}. IEEE,
  \bibinfo{pages}{171--180}.
\newblock


\bibitem[\protect\citeauthoryear{Git}{Git}{2019}]%
        {git}
\bibfield{author}{\bibinfo{person}{Git}.} \bibinfo{year}{2019}\natexlab{}.
\newblock \bibinfo{title}{Git}.
\newblock
\newblock
\urldef\tempurl%
\url{https://git-scm.com}
\showURL{%
\tempurl}
\newblock
\shownote{Last accessed on: 19/09/2019.}


\bibitem[\protect\citeauthoryear{Graves, Karr, Marron, and Siy}{Graves
  et~al\mbox{.}}{2000}]%
        {graves2000predicting}
\bibfield{author}{\bibinfo{person}{Todd~L Graves}, \bibinfo{person}{Alan~F
  Karr}, \bibinfo{person}{James~S Marron}, {and} \bibinfo{person}{Harvey Siy}.}
  \bibinfo{year}{2000}\natexlab{}.
\newblock \showarticletitle{Predicting fault incidence using software change
  history}.
\newblock \bibinfo{journal}{\emph{IEEE Transactions on software engineering}}
  \bibinfo{volume}{26}, \bibinfo{number}{7} (\bibinfo{year}{2000}),
  \bibinfo{pages}{653--661}.
\newblock


\bibitem[\protect\citeauthoryear{Habib and Pradel}{Habib and Pradel}{2018}]%
        {habib2018many}
\bibfield{author}{\bibinfo{person}{Andrew Habib} {and} \bibinfo{person}{Michael
  Pradel}.} \bibinfo{year}{2018}\natexlab{}.
\newblock \showarticletitle{How many of all bugs do we find? a study of static
  bug detectors}. In \bibinfo{booktitle}{\emph{Proceedings of the 33rd ACM/IEEE
  International Conference on Automated Software Engineering}}.
  \bibinfo{pages}{317--328}.
\newblock


\bibitem[\protect\citeauthoryear{Harman, Jia, and Zhang}{Harman
  et~al\mbox{.}}{2015}]%
        {harman2015achievements}
\bibfield{author}{\bibinfo{person}{Mark Harman}, \bibinfo{person}{Yue Jia},
  {and} \bibinfo{person}{Yuanyuan Zhang}.} \bibinfo{year}{2015}\natexlab{}.
\newblock \showarticletitle{Achievements, open problems and challenges for
  search based software testing}. In \bibinfo{booktitle}{\emph{2015 IEEE 8th
  International Conference on Software Testing, Verification and Validation
  (ICST)}}. IEEE, \bibinfo{pages}{1--12}.
\newblock


\bibitem[\protect\citeauthoryear{Hata, Mizuno, and Kikuno}{Hata
  et~al\mbox{.}}{2012}]%
        {hata2012bug}
\bibfield{author}{\bibinfo{person}{Hideaki Hata}, \bibinfo{person}{Osamu
  Mizuno}, {and} \bibinfo{person}{Tohru Kikuno}.}
  \bibinfo{year}{2012}\natexlab{}.
\newblock \showarticletitle{Bug prediction based on fine-grained module
  histories}. In \bibinfo{booktitle}{\emph{2012 34th international conference
  on software engineering (ICSE)}}. IEEE, \bibinfo{pages}{200--210}.
\newblock


\bibitem[\protect\citeauthoryear{Just}{Just}{2019}]%
        {defects4jweb}
\bibfield{author}{\bibinfo{person}{Rene Just}.}
  \bibinfo{year}{2019}\natexlab{}.
\newblock \bibinfo{title}{Defects4J - A Database of Real Faults and an
  Experimental Infrastructure to Enable Controlled Experiments in Software
  Engineering Research}.
\newblock
\newblock
\urldef\tempurl%
\url{https://github.com/rjust/defects4j}
\showURL{%
\tempurl}
\newblock
\shownote{Last accessed on: 02/10/2019.}


\bibitem[\protect\citeauthoryear{Just, Jalali, and Ernst}{Just
  et~al\mbox{.}}{2014}]%
        {just2014defects4j}
\bibfield{author}{\bibinfo{person}{Ren{\'e} Just}, \bibinfo{person}{Darioush
  Jalali}, {and} \bibinfo{person}{Michael~D Ernst}.}
  \bibinfo{year}{2014}\natexlab{}.
\newblock \showarticletitle{Defects4J: A database of existing faults to enable
  controlled testing studies for Java programs}. In
  \bibinfo{booktitle}{\emph{Proceedings of the 2014 International Symposium on
  Software Testing and Analysis}}. ACM, \bibinfo{pages}{437--440}.
\newblock


\bibitem[\protect\citeauthoryear{Kim, Zimmermann, Whitehead~Jr, and Zeller}{Kim
  et~al\mbox{.}}{2007}]%
        {kim2007predicting}
\bibfield{author}{\bibinfo{person}{Sunghun Kim}, \bibinfo{person}{Thomas
  Zimmermann}, \bibinfo{person}{E~James Whitehead~Jr}, {and}
  \bibinfo{person}{Andreas Zeller}.} \bibinfo{year}{2007}\natexlab{}.
\newblock \showarticletitle{Predicting faults from cached history}. In
  \bibinfo{booktitle}{\emph{Proceedings of the 29th international conference on
  Software Engineering}}. IEEE Computer Society, \bibinfo{pages}{489--498}.
\newblock


\bibitem[\protect\citeauthoryear{Le, Lo, and Le~Goues}{Le
  et~al\mbox{.}}{2016}]%
        {le2016history}
\bibfield{author}{\bibinfo{person}{Xuan Bach~D Le}, \bibinfo{person}{David Lo},
  {and} \bibinfo{person}{Claire Le~Goues}.} \bibinfo{year}{2016}\natexlab{}.
\newblock \showarticletitle{History driven program repair}. In
  \bibinfo{booktitle}{\emph{2016 IEEE 23rd International Conference on Software
  Analysis, Evolution, and Reengineering (SANER)}}, Vol.~\bibinfo{volume}{1}.
  IEEE, \bibinfo{pages}{213--224}.
\newblock


\bibitem[\protect\citeauthoryear{Lewis, Lin, Sadowski, Zhu, Ou, and
  Whitehead~Jr}{Lewis et~al\mbox{.}}{2013}]%
        {lewis2013does}
\bibfield{author}{\bibinfo{person}{Chris Lewis}, \bibinfo{person}{Zhongpeng
  Lin}, \bibinfo{person}{Caitlin Sadowski}, \bibinfo{person}{Xiaoyan Zhu},
  \bibinfo{person}{Rong Ou}, {and} \bibinfo{person}{E~James Whitehead~Jr}.}
  \bibinfo{year}{2013}\natexlab{}.
\newblock \showarticletitle{Does bug prediction support human developers?
  findings from a google case study}. In \bibinfo{booktitle}{\emph{Proceedings
  of the 2013 International Conference on Software Engineering}}. IEEE Press,
  \bibinfo{pages}{372--381}.
\newblock


\bibitem[\protect\citeauthoryear{Lewis and Ou}{Lewis and Ou}{2011}]%
        {googledefect}
\bibfield{author}{\bibinfo{person}{Chris Lewis} {and} \bibinfo{person}{Rong
  Ou}.} \bibinfo{year}{2011}\natexlab{}.
\newblock \bibinfo{title}{Bug Prediction at Google}.
\newblock
\newblock
\urldef\tempurl%
\url{http://google-engtools.blogspot.com/2011/12/}
\showURL{%
\tempurl}
\newblock
\shownote{Last accessed on: 16/09/2019.}


\bibitem[\protect\citeauthoryear{Mao, Harman, and Jia}{Mao
  et~al\mbox{.}}{2016}]%
        {mao2016sapienz}
\bibfield{author}{\bibinfo{person}{Ke Mao}, \bibinfo{person}{Mark Harman},
  {and} \bibinfo{person}{Yue Jia}.} \bibinfo{year}{2016}\natexlab{}.
\newblock \showarticletitle{Sapienz: Multi-objective automated testing for
  Android applications}. In \bibinfo{booktitle}{\emph{Proceedings of the 25th
  International Symposium on Software Testing and Analysis}}.
  \bibinfo{pages}{94--105}.
\newblock


\bibitem[\protect\citeauthoryear{Menzies, Greenwald, and Frank}{Menzies
  et~al\mbox{.}}{2006}]%
        {menzies2006data}
\bibfield{author}{\bibinfo{person}{Tim Menzies}, \bibinfo{person}{Jeremy
  Greenwald}, {and} \bibinfo{person}{Art Frank}.}
  \bibinfo{year}{2006}\natexlab{}.
\newblock \showarticletitle{Data mining static code attributes to learn defect
  predictors}.
\newblock \bibinfo{journal}{\emph{IEEE transactions on software engineering}}
  \bibinfo{volume}{33}, \bibinfo{number}{1} (\bibinfo{year}{2006}),
  \bibinfo{pages}{2--13}.
\newblock


\bibitem[\protect\citeauthoryear{Nagappan and Ball}{Nagappan and Ball}{2005}]%
        {nagappan2005use}
\bibfield{author}{\bibinfo{person}{Nachiappan Nagappan} {and}
  \bibinfo{person}{Thomas Ball}.} \bibinfo{year}{2005}\natexlab{}.
\newblock \showarticletitle{Use of relative code churn measures to predict
  system defect density}. In \bibinfo{booktitle}{\emph{Proceedings of the 27th
  international conference on Software engineering}}. ACM,
  \bibinfo{pages}{284--292}.
\newblock


\bibitem[\protect\citeauthoryear{Nagappan, Murphy, and Basili}{Nagappan
  et~al\mbox{.}}{2008}]%
        {nagappan2008influence}
\bibfield{author}{\bibinfo{person}{Nachiappan Nagappan},
  \bibinfo{person}{Brendan Murphy}, {and} \bibinfo{person}{Victor Basili}.}
  \bibinfo{year}{2008}\natexlab{}.
\newblock \showarticletitle{The influence of organizational structure on
  software quality}. In \bibinfo{booktitle}{\emph{2008 ACM/IEEE 30th
  International Conference on Software Engineering}}. IEEE,
  \bibinfo{pages}{521--530}.
\newblock


\bibitem[\protect\citeauthoryear{Nagappan, Zeller, Zimmermann, Herzig, and
  Murphy}{Nagappan et~al\mbox{.}}{2010}]%
        {nagappan2010change}
\bibfield{author}{\bibinfo{person}{Nachiappan Nagappan},
  \bibinfo{person}{Andreas Zeller}, \bibinfo{person}{Thomas Zimmermann},
  \bibinfo{person}{Kim Herzig}, {and} \bibinfo{person}{Brendan Murphy}.}
  \bibinfo{year}{2010}\natexlab{}.
\newblock \showarticletitle{Change bursts as defect predictors}. In
  \bibinfo{booktitle}{\emph{2010 IEEE 21st International Symposium on Software
  Reliability Engineering}}. IEEE, \bibinfo{pages}{309--318}.
\newblock


\bibitem[\protect\citeauthoryear{Oliveira, Aleti, Grunske, and
  Smith-Miles}{Oliveira et~al\mbox{.}}{2018}]%
        {oliveira2018mapping}
\bibfield{author}{\bibinfo{person}{Carlos Oliveira}, \bibinfo{person}{Aldeida
  Aleti}, \bibinfo{person}{Lars Grunske}, {and} \bibinfo{person}{Kate
  Smith-Miles}.} \bibinfo{year}{2018}\natexlab{}.
\newblock \showarticletitle{Mapping the effectiveness of automated test suite
  generation techniques}.
\newblock \bibinfo{journal}{\emph{IEEE Transactions on Reliability}}
  \bibinfo{volume}{67}, \bibinfo{number}{3} (\bibinfo{year}{2018}),
  \bibinfo{pages}{771--785}.
\newblock


\bibitem[\protect\citeauthoryear{Oliveira, Aleti, Li, and Abdelrazek}{Oliveira
  et~al\mbox{.}}{2019}]%
        {aleti19testFunc}
\bibfield{author}{\bibinfo{person}{Carlos Oliveira}, \bibinfo{person}{Aldeida
  Aleti}, \bibinfo{person}{Yuan-Fang Li}, {and} \bibinfo{person}{Mohamed
  Abdelrazek}.} \bibinfo{year}{2019}\natexlab{}.
\newblock \showarticletitle{Footprints of Fitness Functions in Search-Based
  Software Testing}. In \bibinfo{booktitle}{\emph{Proceedings of the Genetic
  and Evolutionary Computation Conference}} \emph{(\bibinfo{series}{GECCO
  ’19})}. \bibinfo{publisher}{Association for Computing Machinery},
  \bibinfo{pages}{1399–1407}.
\newblock
\showISBNx{9781450361118}
\urldef\tempurl%
\url{https://doi.org/10.1145/3321707.3321880}
\showDOI{\tempurl}


\bibitem[\protect\citeauthoryear{Panichella, Kifetew, and Tonella}{Panichella
  et~al\mbox{.}}{2015}]%
        {panichella2015reformulating}
\bibfield{author}{\bibinfo{person}{Annibale Panichella},
  \bibinfo{person}{Fitsum~Meshesha Kifetew}, {and} \bibinfo{person}{Paolo
  Tonella}.} \bibinfo{year}{2015}\natexlab{}.
\newblock \showarticletitle{Reformulating branch coverage as a many-objective
  optimization problem}. In \bibinfo{booktitle}{\emph{2015 IEEE 8th
  international conference on software testing, verification and validation
  (ICST)}}. IEEE, \bibinfo{pages}{1--10}.
\newblock


\bibitem[\protect\citeauthoryear{Panichella, Kifetew, and Tonella}{Panichella
  et~al\mbox{.}}{2017}]%
        {panichella2017automated}
\bibfield{author}{\bibinfo{person}{Annibale Panichella},
  \bibinfo{person}{Fitsum~Meshesha Kifetew}, {and} \bibinfo{person}{Paolo
  Tonella}.} \bibinfo{year}{2017}\natexlab{}.
\newblock \showarticletitle{Automated test case generation as a many-objective
  optimisation problem with dynamic selection of the targets}.
\newblock \bibinfo{journal}{\emph{IEEE Transactions on Software Engineering}}
  \bibinfo{volume}{44}, \bibinfo{number}{2} (\bibinfo{year}{2017}),
  \bibinfo{pages}{122--158}.
\newblock


\bibitem[\protect\citeauthoryear{Panichella, Kifetew, and Tonella}{Panichella
  et~al\mbox{.}}{2018}]%
        {panichella2018large}
\bibfield{author}{\bibinfo{person}{Annibale Panichella},
  \bibinfo{person}{Fitsum~Meshesha Kifetew}, {and} \bibinfo{person}{Paolo
  Tonella}.} \bibinfo{year}{2018}\natexlab{}.
\newblock \showarticletitle{A large scale empirical comparison of
  state-of-the-art search-based test case generators}.
\newblock \bibinfo{journal}{\emph{Information and Software Technology}}
  \bibinfo{volume}{104} (\bibinfo{year}{2018}), \bibinfo{pages}{236--256}.
\newblock


\bibitem[\protect\citeauthoryear{Paterson, Campos, Abreu, Kapfhammer, Fraser,
  and McMinn}{Paterson et~al\mbox{.}}{2019}]%
        {paterson2019empirical}
\bibfield{author}{\bibinfo{person}{David Paterson}, \bibinfo{person}{Jose
  Campos}, \bibinfo{person}{Rui Abreu}, \bibinfo{person}{Gregory~M Kapfhammer},
  \bibinfo{person}{Gordon Fraser}, {and} \bibinfo{person}{Phil McMinn}.}
  \bibinfo{year}{2019}\natexlab{}.
\newblock \showarticletitle{An Empirical Study on the Use of Defect Prediction
  for Test Case Prioritization}. In \bibinfo{booktitle}{\emph{2019 12th IEEE
  Conference on Software Testing, Validation and Verification (ICST)}}. IEEE,
  \bibinfo{pages}{346--357}.
\newblock


\bibitem[\protect\citeauthoryear{Pearson, Campos, Just, Fraser, Abreu, Ernst,
  Pang, and Keller}{Pearson et~al\mbox{.}}{2017}]%
        {pearson2017evaluating}
\bibfield{author}{\bibinfo{person}{Spencer Pearson}, \bibinfo{person}{Jos{\'e}
  Campos}, \bibinfo{person}{Ren{\'e} Just}, \bibinfo{person}{Gordon Fraser},
  \bibinfo{person}{Rui Abreu}, \bibinfo{person}{Michael~D Ernst},
  \bibinfo{person}{Deric Pang}, {and} \bibinfo{person}{Benjamin Keller}.}
  \bibinfo{year}{2017}\natexlab{}.
\newblock \showarticletitle{Evaluating and improving fault localization}. In
  \bibinfo{booktitle}{\emph{Proceedings of the 39th International Conference on
  Software Engineering}}. IEEE Press, \bibinfo{pages}{609--620}.
\newblock


\bibitem[\protect\citeauthoryear{Rahman, Posnett, Hindle, Barr, and
  Devanbu}{Rahman et~al\mbox{.}}{2011}]%
        {rahman2011bugcache}
\bibfield{author}{\bibinfo{person}{Foyzur Rahman}, \bibinfo{person}{Daryl
  Posnett}, \bibinfo{person}{Abram Hindle}, \bibinfo{person}{Earl Barr}, {and}
  \bibinfo{person}{Premkumar Devanbu}.} \bibinfo{year}{2011}\natexlab{}.
\newblock \showarticletitle{BugCache for inspections: hit or miss?}. In
  \bibinfo{booktitle}{\emph{Proceedings of the 19th ACM SIGSOFT symposium and
  the 13th European conference on Foundations of software engineering}}. ACM,
  \bibinfo{pages}{322--331}.
\newblock


\bibitem[\protect\citeauthoryear{Rojas, Vivanti, Arcuri, and Fraser}{Rojas
  et~al\mbox{.}}{2017}]%
        {rojas2017detailed}
\bibfield{author}{\bibinfo{person}{Jos{\'e}~Miguel Rojas},
  \bibinfo{person}{Mattia Vivanti}, \bibinfo{person}{Andrea Arcuri}, {and}
  \bibinfo{person}{Gordon Fraser}.} \bibinfo{year}{2017}\natexlab{}.
\newblock \showarticletitle{A detailed investigation of the effectiveness of
  whole test suite generation}.
\newblock \bibinfo{journal}{\emph{Empirical Software Engineering}}
  \bibinfo{volume}{22}, \bibinfo{number}{2} (\bibinfo{year}{2017}),
  \bibinfo{pages}{852--893}.
\newblock


\bibitem[\protect\citeauthoryear{Rueda, Vos, and Prasetya}{Rueda
  et~al\mbox{.}}{2015}]%
        {rueda2015unit}
\bibfield{author}{\bibinfo{person}{Urko Rueda}, \bibinfo{person}{Tanja~EJ Vos},
  {and} \bibinfo{person}{ISWB Prasetya}.} \bibinfo{year}{2015}\natexlab{}.
\newblock \showarticletitle{Unit Testing Tool Competition--Round Three}. In
  \bibinfo{booktitle}{\emph{2015 IEEE/ACM 8th International Workshop on
  Search-Based Software Testing}}. IEEE, \bibinfo{pages}{19--24}.
\newblock


\bibitem[\protect\citeauthoryear{Shamshiri, Just, Rojas, Fraser, McMinn, and
  Arcuri}{Shamshiri et~al\mbox{.}}{2015}]%
        {shamshiri2015automatically}
\bibfield{author}{\bibinfo{person}{Sina Shamshiri}, \bibinfo{person}{Rene
  Just}, \bibinfo{person}{Jose~Miguel Rojas}, \bibinfo{person}{Gordon Fraser},
  \bibinfo{person}{Phil McMinn}, {and} \bibinfo{person}{Andrea Arcuri}.}
  \bibinfo{year}{2015}\natexlab{}.
\newblock \showarticletitle{Do automatically generated unit tests find real
  faults? an empirical study of effectiveness and challenges (t)}. In
  \bibinfo{booktitle}{\emph{2015 30th IEEE/ACM International Conference on
  Automated Software Engineering (ASE)}}. IEEE, \bibinfo{pages}{201--211}.
\newblock


\bibitem[\protect\citeauthoryear{Vargha and Delaney}{Vargha and
  Delaney}{2000}]%
        {vargha2000critique}
\bibfield{author}{\bibinfo{person}{Andr{\'a}s Vargha} {and}
  \bibinfo{person}{Harold~D Delaney}.} \bibinfo{year}{2000}\natexlab{}.
\newblock \showarticletitle{A critique and improvement of the CL common
  language effect size statistics of McGraw and Wong}.
\newblock \bibinfo{journal}{\emph{Journal of Educational and Behavioral
  Statistics}} \bibinfo{volume}{25}, \bibinfo{number}{2}
  (\bibinfo{year}{2000}), \bibinfo{pages}{101--132}.
\newblock


\bibitem[\protect\citeauthoryear{Zimmermann, Premraj, and Zeller}{Zimmermann
  et~al\mbox{.}}{2007}]%
        {zimmermann2007predicting}
\bibfield{author}{\bibinfo{person}{Thomas Zimmermann}, \bibinfo{person}{Rahul
  Premraj}, {and} \bibinfo{person}{Andreas Zeller}.}
  \bibinfo{year}{2007}\natexlab{}.
\newblock \showarticletitle{Predicting defects for eclipse}. In
  \bibinfo{booktitle}{\emph{Third International Workshop on Predictor Models in
  Software Engineering (PROMISE'07: ICSE Workshops 2007)}}. IEEE,
  \bibinfo{pages}{9--9}.
\newblock


\end{thebibliography}

\end{NoHyper}
\end{document}